\documentstyle[12pt,titlepage,psfig]{article}
\begin{document}

\pagestyle{empty}
\title{An Eulerian PPM \& PIC Code for Cosmological Hydrodynamics}
\author{Andrew Sornborger,\\
Cambridge University, DAMTP, \\
Silver Street, Cambridge CB3 9EW, UK\\
${\rm Bruce \: Fryxell^{1}}$,
${\rm Kevin \: Olson^{1}}$,\\
Institute for Computational Science and Informatics,\\
George Mason University, Fairfax VA 22030 \\
and \\
${\rm Peter \: MacNeice^{1}}$\\
Hughes STX, Greenbelt, MD 20770\\
\\
\\
{\footnotesize 1) Postal Address: NASA GSFC, Code 934, Greenbelt, MD 20771}
}
\date{}
\maketitle


\pagestyle{myheadings}
\markright{  }
\newpage
\begin{center}
\bf
Abstract
\end{center}
\rm 
We present a method for integrating the cosmological
hydrodynamical equations including a collisionless dark matter
component. For modeling the baryonic matter component, we use the
Piecewise Parabolic Method (PPM) which is a high-accuracy shock
capturing technique. The dark matter component is modeled using
gravitationally interacting particles whose evolution is determined
using standard particle-in-cell techniques. We discuss details of the
inclusion of gravity and expansion in the PPM code and give results of
a number of tests of the code.  This code has been developed for a
massively parallel, SIMD supercomputer: the MasPar MP-2 parallel
processor. We present details of the techniques we have used to
implement the code for this architecture and discuss performance of
the code on the MP-2. The code processes $5.0 \times 10^4$ grid zones
per second and requires 53 seconds of machine time for a single
timestep in a $128^3$ simulation.\\ 
{\em Subject Headings:} methods: numerical

\newpage

\section{Introduction}

To study the matter distribution in the universe on scales less than
$\sim 5 Mpc$, it is necessary to take into account the contribution of
both baryonic and dark matter. Baryonic matter has a small mean free
path on cosmological scales and is therefore treated as a compressible
fluid. The dark matter is assumed to be collisionless and to
contribute only to the overall gravitational field (although some
cosmological models without collisionless matter have been proposed,
one of the authors (A.S.) is using the code to study structure
formation in the cosmic string model, which does assume a dark matter
component, as do many other cosmological models). We have developed a
computational code to simulate a collisional (baryonic) and
collisionless (dark matter) fluid together, based on two numerical
techniques: the Piecewise Parabolic Method (PPM) method (Colella \&
Woodward 1984, Woodward \& Colella 1984) for the integration of the
equations describing the compressible, baryonic fluid and the
particle-in-cell (PIC) method (Hockney \& Eastwood 1988) for
integration of the evolution equations describing the collisionless
dark matter. In this paper, we use `PIC' as opposed to the more common
appellation `PM' to avoid confusion with other common abbreviations
used for other numerical techniques (e.g. PPM for piecewise parabolic
method).

A code similar to ours has also been developed (Bryan et al. 199(?))
which uses a Lagrangian step plus remap technique also outlined in
Colella \& Woodward (1984). The group that developed the code went to
some lengths to include good resolution at length scales of 2-3 grid
spacings by introducing corrections to the Riemann solver and to
accurately track pressures using a dual tracking method.  One of us
(B.F.) along with P. Ricker at Chicago is currently developing a PPM
based cosmological hydrodynamical code which uses the dual tracking
method to accurately track pressures. Accurate pressure tracking is
crucial for codes which are designed to investigate the effects of
ionization and recombination in astrophysical processes.  Currently,
our code does not include such techniques since we intend to apply the
code to the simulation of the dynamics of matter on large scales,
neglecting ionization and other such pressure dependent processes.

\subsection{Equations}

From the isotropic big bang cosmological model, we take the fluids to
exist in an expanding Friedmann -- Robertson -- Walker -- Lema\^itre
background with scale factor $a(t)$ giving the distance scale at time
$t$. The expansion rate is then given by $H \equiv \frac{\dot a}{a}$. 

The baryonic fluid equations are given by the covariant divergence of
the stress-energy tensor, and gravity obeys the Einstein equations. For
non-relativistic velocities, and pressures much less than the rest
mass of the fluid particles, the Euler equations, which govern the
behavior of baryonic matter take the form
\begin{equation}
   \partial_t \rho + \partial_k (\rho v_k) = 0
\end{equation}
\begin{equation}
   \partial_t (\rho v_i) + \partial_k (\rho v_k v_i + \delta_{ij} p) =
      -2 \frac{\dot a}{a} \rho v_i - \frac{\rho}{a^3} \partial_i \phi
\end{equation}
\begin{equation}
   \partial_t (\rho E) + \partial_k (\rho E + p) v_k = -4 \frac{\dot
      a}{a} \rho E - \frac{\rho}{a^3} v_k \partial_k \phi 
\end{equation}

The above equations assume variables related to physical variables as
follows: $\rho \equiv a^3 \rho_p$, $p \equiv a p_p$, $a^2 u \equiv
u_p$, $a^2 T \equiv T_p$, $a^2 E \equiv E_p$, and $\phi \equiv a
\phi_p + \frac{a^2 \ddot a}{2} x_p$. 

Here $\rho$ is density, $p$ is pressure, $u$ is internal energy, $T$
is temperature, $E = \frac{1}{2} v^2 + u$ is total specific energy,
and $v$ is velocity. Comoving spatial coordinates are used where $x_p
\equiv a x$ and $\dot x_p \equiv v_p = \dot a x + a \dot x \equiv \dot
a x + a v$. Variables underscored with $p$ are physical variables.

These equations are equivalent to those given by Peebles (1980) and
Cen (1992) except that the variables are such that the differential
operators on the left hand sides are the same as the nonexpanding
Euler equations as in Bryan, et. al. (199(?)). This helps make the
implementation of the PPM method relatively straightforward.

We also assume an ideal gas equation of state,
\begin{equation}
   \rho u (\gamma - 1) = p,
\end{equation}
and an adiabatic gas,
\begin{equation}
   p = c\rho T,
\end{equation}
One can check that these relations remain unchanged under the change
from physical to the above defined variables.

Dark matter obeys the collisionless form of the Boltzmann equation,
called the Vlasov equation. Writing first order equations in
Lagrangian variables, with the above definitions for velocity $v$,
density $\rho$, scale factor $a(t)$ and potential $\phi$ we find
equations for dark matter particles,
\begin{equation}
   \dot v_i + \frac{2 \dot a}{a} v_i = -\frac{1}{a^3} \partial_i \phi,
\end{equation}
\begin{equation}
   \dot r_i = v_i.
\end{equation}

In the Newtonian approximation to the Einstein equations the Poisson
equation gives the gravitational potential $\partial^2 \phi = 4 \pi G
(\rho_{tot} - \bar\rho_{tot})$ where $\bar \rho$ is the average
background density and is equal to $3 \ddot a a^{2}$. The subscript
$tot$ indicates that the densities here are the total matter densities
and are equal to the sum of the baryonic and the dark matter
densities.

\section{Integration Techniques}

The driving issue in selection of techniques for a cosmological
hydrodynamical code is accurate resolution of non-linear effects. In
particular, we want good shock resolution since shocks are ubiquitous
in cosmological flows. PPM is a method which has been well tested as
an accurate method for treating flows with discontinuities.  Since PPM
is grid based, it is most natural to use a grid based method to model
the dark matter distribution as well. PIC is an extensively tested,
partially particle based, partially grid based method (Hockney \&
Eastwood 1985) which we have combined with PPM to form our collisional
plus collisionless fluid code.

\subsection{The Piecewise Parabolic Method}

PPM is a higher order Godunov method for integrating partial
differential equations (Colella \& Woodward 1984, Woodard \& Colella
1984). The code which we built upon to make a cosmological code has
been tested in many highly non-linear astrophysical fluid
scenarios. It was originally developed to study the dynamics of
supernova explosions (Fryxell, Muller, \& Arnett 1991) and as such
includes a Riemann solver which is capable of treating non-gamma law
gases.  Since the cosmological fluid equations given above assume an
ideal fluid equation of state, the sophistication is only of use for
flat space simulations.

\subsubsection{The Godunov Method}

The Godunov method is a finite volume method. This means that the
fluid equations are considered in integral form and thus the problem
of calculating divergences becomes a problem of calculating fluxes and
thus mass, momentum and energy are exactly conserved, barring the
introduction of source terms such as the bulk expansion terms in the
cosmological Euler equations given above.

In a finite volume method, one divides the simulation volume into a
set of zones (sometimes called cells), each of which contains values
corresponding to the total mass, velocity or energy (as well as other
quantities) in that volume. One then uses the integral Euler equations
to find a solution. For instance, the continuity equation
\begin{equation}
   \int d^3x \partial_t \rho + \int d^3x \vec \nabla \cdot \rho \vec v
      = 0 
\end{equation}
becomes
\begin{equation}
   \partial_t \bar \rho + \sum_{sides} \rho \vec v \cdot \vec S = 0
\end{equation}
where $\bar \rho$ means the total density in the gridzone and $\vec S$
is a normal vector for a given side of the volume.

To determine the time evolution of these quantities, one must
determine the fluxes to and from the gridzones over a (small) fixed
time interval determined, typically, by the Courant condition. The
Courant condition determines the maximum time that one can integrate
and still maintain causality in the integration.

The Godunov method uses the approximation that the quantities within
each zone are spatially flat. Therefore, for instance the sound speed
is considered to be constant throughout the entire volume, as are all
other fluid variables. This assumption is the first step in the
Godunov method.

The second step is the physical step. To determine the fluxes from one
gridzone to the next, one solves the Riemann shock tube problem
(exactly, if no source terms are present) at zone interfaces. The
solution to the Riemann problem assumes that initially, the states on
either side of an interface are spatially constant. The solution to
the Riemann problem is a set of non-linear discontinuities in the
state variables propagating from each interface with characteristic
velocities. Using these propagating discontinuities at the interface
one can calculate the difference between the initial state and the
solution after a given time interval and thus find the fluxes from and
to each zone which are then used in the third step, in which state
variable averages are updated. Once new averages are obtained, the
successive timestep is calculated. 

The advantage of the Godunov method is that non-linearity is
introduced into the differencing scheme via solution of the Riemann
problem. Linear schemes for calculating fluxes force one to choose
between the width of a discontinuity and the amplitude of oscillations
propagating away from the discontinuity due to the Gibbs effect.
Linear schemes also spuriously allow sound waves to propagate upwind
in supersonic flows. Both of these effects are avoided in the Godunov
method.

\subsubsection{PPM}

PPM introduces a number of changes to achieve higher order resolution
in a Godunov method. The states for input to the Riemann solver are
still assumed to be spatially constant, but better accuracy is
obtained in the evaluation of average quantities within the causal
radius of the zone interfaces by the introduction of interpolated
parabolae. Thus, instead of using a flat contour, as in the simplest
Godunov method, which contains no information on subzone scales, one
uses a higher order contour to get better spatial information within
each zone. The spatial information from the parabolae is used to
better determine the initial data for input to the Riemann solver. The
way this works is one makes a guess at the spatially adjacent left and
right states for input to the Riemann solver. Then the guess is
corrected using the linearized characteristic equations. Using the
corrected characteristic speeds, averages are taken over the causal
regions of the interpolated parabolae; With some corrections to include
the effects of body forces and to insure higher order accuracy, the
averages are used as the states for input to the Riemann solver.

To dampen oscillations at shocks, the parabolae are required to be
monotonic, and flattening is introduced near shocks to damp the
oscillations. Due to the introduction of these constraints, artificial
numerical viscosity, which must be introduced to further dampen the
oscillations, can be kept at levels much less than most other
techniques for integrating the fluid equations.  As a result,
discontinuities which are one to two grid points wide can be followed
without generating significant unphysical oscillations. For details,
the reader should refer to the original paper by Colella and Woodward
(1984).

\subsection{Changes to PPM due to Gravity}

We have used what we feel are the minimal changes necessary to
introduce the gravitational and expansion terms to the PPM code. The
inclusion of source terms is outlined in Colella and Woodward
(1984). There are two areas where changes are necessary:
the states input to the Riemann solver must include corrections due to
gravity (expansion is a homogeneous term and thus does not contribute
to gradient effects); and the gravity and expansion terms must be
added to the update step.

The gravitational potential is first calculated at each grid point
using a standard FFT Poisson solver as outlined in Hockney \& Eastwood
(1988).  To implement the corrections for the Riemann solver states,
we interpolate parabolae for the gravitational force at timestep $n$,
then use values for the gravitational force at the zone interface to
calculate the solution to the modified Riemann problem. See Colella
and Woodward (1984) p. 191.

The PPM update step (see C\&W p. 191), when one includes gravity and
expansion source terms, requires that we know the values $\rho^{n +
\frac{1}{2}}$, $v^{n + \frac{1}{2}}$ and $E^{n + \frac{1}{2}}$. This
renders the code implicit. We need to overcome this problem while
retaining second order accuracy in the code. This can be done by
calculating approximate values $\rho'^{n + 1}$, $v'^{n + 1}$ and
$E'^{n+1}$ for the equations with no source terms. Then we use the
average $\rho^{n + \frac{1}{2}} = \frac{1}{2} (\rho^n + \rho'^{n +
1})$ as input to the Poisson solver to calculate $g^{n +
\frac{1}{2}}$. And we use the averages $v^{n + \frac{1}{2}} =
\frac{1}{2} (v^n + v'^{n + 1})$ and $E^{n + \frac{1}{2}} = \frac{1}{2}
(E^n + E'^{n + 1})$ plus the gravitational force to update the state
variable averages.

\subsection{The Particle-in-Cell Method}

The particle-in-cell method (Hockney \& Eastwood 1985) uses particles
to statistically represent mass density in a collisionless fluid. The
method consists of five steps: 
\begin{enumerate}
\item The particle masses are deposited via interpolation onto a mesh
to give the mass density as a function of position.
\item The gravitational potential is calculated by solving Poisson's
equation on the mesh (we use the same FFT Poisson solver for this step
as for the PPM integration).
\item Forces are calculated by finite differencing of the potential on
the mesh.
\item These forces are then interpolated back to the particle
locations.
\item The particle positions and velocities are updated using the
Lagrangian equations of motion.
\end{enumerate}
To perform the interpolation steps mentioned above we use
cloud-in-cell interpolation. That is, a particle contributes a 
fraction of its mass to each of its 8 surrounding mesh cells which
varies linearly with the particles' relative position measured with
respect to that mesh cell.

\subsection{Combining PPM and PIC}

Since baryonic matter and dark matter interact only gravitationally,
combination of the codes is straightforward (but not trivial) since
the fluids only interact via combination of their gravitational
potentials. The main consideration in combining the codes is that the
integration steps are slightly different. Eulerian PPM updates
variables in two operator splitting sweeps of equal integration time
$dt$, this allows the code to remain second order in time. The PIC
code can in principle change integration time $dt$ at each step.

At the beginning of a timestep, we have dark matter particle positions
$x_i$ at timestep $n - \frac{1}{2}$, dark matter particle velocities
$v_i$ at timestep $n$; and baryonic fluid variables $\rho$, $v$ and
$E$ are defined at timestep $n$.

We combine the integrations as follows: 
\begin{enumerate}
\item At the beginning of the timestep for the first operator
splitting sweep, we advance the particle positions to timestep $n +
\frac{1}{2}$ using the previous timestep $dt$ and the previous
gravitational potential from time level $n-\frac{1}{2}$.  We use the
particle velocities, along with the fluid velocity at timestep $n$ to
calculate a new timestep $dt_{new}$.
\item Using the new timestep $dt_{new}$, we correct the particle
positions at timestep $n + \frac{1}{2}$ to be centered for the new
timestep $(n + \frac{1}{2})'$.
\item We update the fluid variables not including corrections for
gravitation or expansion to get uncorrected values for $\rho$, $v$ and
$E$ at timestep $n + 1$. For the first sweep, we calculate first x,
then y, then z fluxes.
\item We estimate the fluid density at timestep $n + \frac{1}{2}$ by
averaging the uncorrected value for $\rho$ at timestep $n + 1$
obtained in step 3. with $\rho$ at timestep $n$. This density is then
added to the dark matter density to obtain the total density at each
mesh location.  Steps 3. and 4. are explained above in section 2.2.
\item We use the total density to calculate the gravitational
potential using the FFT Poisson solver. Then using finite differences,
we obtain the gravitational force and use it to update the particle
velocities to time step $n + 1$.
\item The fluid state variables are then corrected with the
gravitational and expansion terms and fluid variables are obtained at
timestep $n + 1$.
\end{enumerate}

For the second operator splitting sweep, we proceed as above reversing
the order of the flux calculation (to z, then y, then x) but with
fixed timestep (reusing $dt_{new}$). This process is repeated for each
2 time steps over the course of a complete simulation.

\subsection{Timescales}

There are three physically relevant timescales in cosmological
hydrodynamics: the expansion rate, the fluid velocity timescale and
the gravitational freefall timescale.

For accurate integration of the expansion source terms, we require
that the simulation volume not expand more than $1\%$ per timestep.
This implies a timestep $\Delta t < \frac{1}{100 H}$, where $H =
\frac{\dot a}{a}$.We also constrain the timestep such that no
information can travel more than a fraction of a zone (typically
$30\%$) in a single timestep. This constraint is also applied to
particles in the dark matter simulation. This gives the constraint
$\Delta t < 0.3 \frac{\Delta x}{|\vec v_{max}|}$.  We further
constrain the timestep to be less than the free-fall time estimated
from the maximum density $\Delta t < 0.3 \frac{1}{\sqrt{\frac{4 \pi G
\rho}{a}}}$.  Finally, we keep the timestep from changing by more than
$25\%$ from timestep to timestep.

\section{Implementation on the MasPar MP-2}

In this section, we describe the implementation of the PPM \& PIC code
on a MasPar MP-2 parallel processor. The MP-2 is an ``inexpensive''
parallel processor which is efficient for grid based integration
methods due to the grid-like nature of its processor layout and
its efficient near neighbor communications network.

\subsection{The Maspar MP-2 Architecture}

The MasPar MP-2 at Goddard Space Flight Center has a SIMD architecture
with 16384 processors. The nominal peak performance is 6.2Gflops.
Each processor has 64Kb of dedicated data memory. The processors are
arranged in a 2D array with dimensions $128\times 128$.  Straightline
connections, known collectively as the X-net, exist between processors
in the north, south, east, west, north-east, south-east, south-west
and north-west directions. At the edges of the processor array the
X-net wraps around so that the array has the same topology as the
surface of a torus.  Inter-processor communications can be achieved in
one of two ways. The global router can be used for more complex
patterns or for communication between widely spaced processors, while
for regular patterns over short distances the X-net communications are
much more efficient.\footnote{A plural floating point multiply takes
40 clocks on the MP-2, an X-net operation sending a real number a
distance of 1 processor takes 41 clocks, and a random communication
pattern using the global router, with all processors participating
takes $\sim 5000$ clocks.}  The MasPar series broadens the definition
of SIMD in at least one important way. It enables indirect addressing
within a processor memory.

As is apparent, the Xnet communications of the Maspar MP-2 are
particularly useful for grid based numerical techniques since the
numerical grid can easily be mapped to the processors and information
from adjacent or nearly adjacent processors is passed very
quickly. Thus, finite differences can be computed efficiently in
parallel.  Further, we wish to keep the use of the global router
to a minimum.

The code we describe here was implemented in Maspar Fortran which is
a subset (plus extensions) of the Fortran 90 standard.

\subsection{Implementing the PPM Code}

The PPM code implementation on the MP-2 uses $3D$ arrays of state
variables to store the fluid state at each timestep. One dimension of
each $3D$ array was stored in processor memory and two dimensions were
distributed across the processor grid.  Thus, each processor contains a
column of mesh points.

As mentioned above our code uses a sophisticated Riemann solver which
calculates the fluid equation of state which can vary from cell to
cell. This is useful, for instance, when investigating stellar
interiors. The number of temporary arrays required by this and other
parts of the PPM algorithm was too large for calculations to be
carried out in $3$-dimensional arrays. Therefore, we adopted the
method of swapping $2D$ subarrays into scratch arrays with dimensions
such that the data is distributed across the processor grid.  The
fluxes are then computed within these scratch arrays.  This technique
also allows the compiler to take better advantage of the processor
registers and generate more efficient code.

Using operator splitting, we calculate the flux in a given dimension
by successively swapping all subarrays in a given direction and
calculating the fluxes for that dimension.  The same is then done for
each of the other two dimensions (for a $3$-dimensional
calculation). Thus, for a simulation with $N^{3}$ mesh cells, we do $N
\times 3$ swaps for each complete sweep.

Extra time is required to swap subarrays with one dimension stored in
processor memory to a dimension distributed across the processors.
Currently, we are swapping subarrays one at a time, but are updating
the code to swap the entire array at once which should save a
considerable amount of time.

For each timestep of the PPM algorithm, the breakdown of calculation
times for a $128^3$ volume is as follows: each flux calculation (there
are three) takes roughly $8$ seconds of cpu time and the swaps of
subarrays from and to the storage arrays also take roughly $8$
seconds (total of $16$ seconds). The FFT Poisson solver uses
efficiently implemented MasPar library functions and takes $2.2$
seconds to execute. Thus, an entire timestep requires $42$ seconds of
machine time per $128^3$ simulation volume, this is equivalent to a
throughput of $5.0 \times 10^4$ gridzones per second. 

We think that replacing the sophisticated Riemann solver with the
solver used by Colella and Woodward (1984) will cut the run time in
half; it will also cut down the number of $3D$ arrays required,
potentially leaving more memory for larger simulations.  Therefore, we
plan to make this change in the near future.

\subsection{Implementing the PIC Code}

We spent considerable effort to efficiently implement the PIC code on
the Maspar MP-2. While the cost of the computation of the
gravitational potential and its finite differences on the mesh is
extremely fast since we have used the highly tuned FFT routine
supplied by Maspar, other parts of the PIC algorithm are more
difficult to parallelize.  These steps are the interpolation of the
particle data to the mesh and the subsequent interpolations of the
forces computed on the mesh back to the particles.  While these steps
generally comprise a small fraction of the cost of the overall
algorithm on serial machines, they constitute the bulk of the running
time of the algorithm on a fine grained parallel machine due to the
fact that two different data structures which are laid out
differently on the processor array, must communicate: namely the
particle list and the computational mesh.  Nonetheless it is possible
to parallelize these steps.  Here, we briefly describe various methods
for the parallelization of the entire PIC algorithm paying particular
attention to the interpolation steps (see MacNeice, Mobarry, \& Olson
1995) for details), then we compare the methods and show under which
circumstances the various methods are efficient and why we choose the
method we use.

To parallelize the PIC code we have to map both an algorithm and a
data structure to the architecture.  The four basic steps in a PIC
algorithm are 
\begin{enumerate}
\item Interpolate the particle data to the computational mesh and
compute the mass density on the mesh. This step is a scatter with add. The
model particles are small but finite sized charge clouds which
contribute to the mass density of any grid cells with which they
overlap.  We use cloud-in-cell interpolation so that a particle will
contribute mass to at most 8 mesh cells.
\item Solve for $\phi$ and then the force $\vec g$ at the grid points.
\item Interpolate $\vec g$ to the particle locations in order to
estimate the force acting on each particle. This is a gather step. 
\item Push the particles, ie. integrate the equations of motion over
$\Delta t$ for each particle.
\end{enumerate}
In combination these four steps involve computation and communication
between two different data structures. The field data has the
qualities of an ordered array in the sense that each element has
specific neighbors. The particle data has the qualities of a randomly
ordered vector, in which element $i$ refers to particle $i$, and no
element has any special relationship to its neighbors in the vector.

Steps 2 and 4 are parallelizable in rather obvious ways, since they
involve only simple and completely predictable data dependencies, and
do not couple the two data structures. Steps 1 and 3 however do couple
the two data structures, with complicated and unpredictable data
dependencies which evolve during the simulation. It is these steps
which invariably dominate the execution times of parallel PIC codes.

On a serial machine the PIC code will execute its computational
workload in a time which is independent of any correlations in the
spatial locations of the particles. This is not true on parallel
machines, such as the MasPar. Spatial clustering of particles can
create communication and/or computational hot-spots which impair
performance.


\subsubsection{Deposition of Mass}

The algorithm we opt to use for the interpolation of particle
information (i.e. mass) to the computational mesh in the code assumes
that the particle data is distributed evenly across the processor
array paying no regard to the physical location of the particles.
This ensures computational load balance in the particle push step.
The scatter-with-add of step 1 is performed using the global router to
perform a sendwithadd.  Maspar Fortran provides a compiler directive
(known as the `collisions' directive) which can be inserted in the
code at the appropriate location so that this function is performed.
The collisions directive handles message contention at a receiving
processor by accepting one of the messages being sent to it and
instructing the rest to try again.  Eventually all are received and
are successfully accumulated in the mass density array.  Clearly the
execution time for this algorithm is set by the processor which has to
receive the most messages, and so this scheme will suffer
communication hotspots in the event of spatial clustering of
particles.

Other algorithms have been tested for this problem and we refer the
interested reader to MacNeice et al. (1995) for a discussion of the
details of the performance charateristics of these techinques.  We
have opted for the above described technique since it was by far the
easiest to implement and its use does not significantly impact the
overall running time of the complete algorithm.

The breakdown of the timing for the PIC portion of the code is as
follows: For a simulation with $128^3$ particles on a grid with
$128^3$ gridzones, the deposit takes about $1.4$ seconds of machine
time. The FFT Poisson solver takes $2.2$ seconds. The force
interpolation takes $3.3$ seconds. Thus, a complete timestep requires
about $6.9$ seconds of machine time.

\section{Code Tests}

The PPM code we use has been tested extensively on a number of
problems for the case with no source terms present. See Fryxell,
Muller, \& Arnett 1991, Fryxell, Zylstra, \& Melia 1992, and Fryxell
\& Taam 1989 for a representative presentation of tests and
results. Below, we present results for the code including expansion
and gravitational source terms.

\subsection{Testing PPM with Gravity}

We have made five tests of the PPM code described above to check that
it is solving the cosmological fluid equations correctly. The first
three tests are actual comparisons to solutions of the equations. The
fourth test checks how the solution degrades as a function of 
resolution, and the fifth test checks that the solutions to one
dimensional initial conditions relax to self-similar solutions (as
they should in an $\Omega = 1$ universe due to the lack of a
characteristic length scale in the equations).

\subsection{Homogeneous Expansion}

In this test, we introduce a bulk velocity and temperature to the
fluid, but the initial conditions are spatially constant. This gives
the equations
\begin{equation}
   \partial_t \rho = 0
\end{equation}
\begin{equation}
   \partial_t (\rho v) = -\frac{2 \dot a}{a} \rho v
\end{equation}
\begin{equation}
   \partial_t (\rho E) = -\frac{4 \dot a}{a} \rho E
\end{equation}
which have solutions where $\rho$ remains constant, $\rho v$ goes as
$a^{-2}$ and $\rho E$ goes as $a^{-4}$.

The results from the code are plotted in figures 1 and 2. In both
cases, the numerical results match the analytical results to a
fraction of one percent. Remember that the scale factor $a(t) =
(\frac{t}{t_0})^{\frac{2}{3}}$, thus the amount of expansion for a
given simulation may be calculated as $\frac{a_{final}}{a_{init}} =
(\frac{t_{final}}{t_{init}})^{\frac{2}{3}}$.

\subsection{Non-Expanding Jeans Length}

The non-expanding Jeans length test is a test of gravitational and
pressure forces with no expansion. We start with the mass and momentum
conservation equations in $1$-dimension:
\begin{equation}
   \partial_t \rho + \partial (\rho v) = 0
\end{equation}
\begin{equation}
   \partial_t v + v \partial v = -\frac{1}{\rho}\partial p - g
\end{equation}
\begin{equation}
   \partial^2 \phi = 4 \pi G \rho
\end{equation}
where $\partial \equiv \partial_x$ and $v \equiv v_x$. We can obtain a
solution for small perturbations by linearizing around background
values 
\begin{equation}
   \rho = \rho_0 + \epsilon\rho_1, p = p_0 + \epsilon p_1, v =
      \epsilon v_1, g = \epsilon g_1
\end{equation}
To first order in $\epsilon$ and combining the equations we find
\begin{equation}
   \partial_t^2 \rho_1 - [v_s^2 \partial^2 \rho_1 + 4 \pi G \rho_1
      \rho_0] = 0
\end{equation}
where $v_s \equiv \frac{\partial p}{\partial \rho} = \frac{\gamma
p}{\rho}$. And expanding in spatial fourier modes we find, for the
time evolution of the amplitudes
\begin{equation}
   \ddot \rho_1 + [v_s^2 k^2 - 4 \pi G \rho_0]\rho_1 = 0
\end{equation}
We see that there exists a critical pressure at $v_s^2 k^2 = 4 \pi G
\rho_0$, where $k$ is the wavenumber, where the gravitational and
pressure forces balance. For pressures above the critical pressure
$p_{crit} = \frac{4 \pi G \rho_0^2}{\gamma k^2}$ we obtain wavelike
solutions
\begin{equation}
   \rho = \rho_0 (1 + \epsilon \cos{kx + \omega t})
\end{equation}
\begin{equation}
   p = p_0 (1 + \epsilon \gamma \cos{kx + \omega t})
\end{equation}
\begin{equation}
   v = -\epsilon \frac{\omega}{k} \sin{kx + \omega t}
\end{equation}
Below the critical pressure we find collapsing solutions with
amplitude increasing exponentially in time. These solutions are the
same as above but with $\omega$ continued to $i \omega$.

To test this perturbative solution, we set up a small amplitude
sinusoidal perturbation given by the above solution and allow it to
evolve while making sure that the density amplitude does not exceed a
small fraction of the background density. We then fourier decompose
the resulting density and compare the evolution of the amplitude of
the first fourier mode (the mode with wavelength equal to the volume
size) to the analytical solution.

The evolution of the first fourier mode obtained from the simulation
code for the wavelike case is given in figure 3 and the result from
the collapsing case is given in figure 4.

In the figures, the analytical solution is plotted against the
amplitude of the first fourier mode. Note that in the collapsing case
the collapse fails to match the exponential expansion given by the
analytical solution. This is because the matter becomes concentrated
in one grid zone and, because of finite resolution, the density cannot
increase further. In the wavelike case, the cosine and sine modes
(both modes with wavelength the same as the size of the volume)
oscillate as the wave moves across the simulation volume, giving the
oscillating amplitudes shown in the figure, but the combined amplitude
remains constant, as it should.

\subsection{Expanding Jeans Length}

The expanding Jeans length test includes expansion along with gravity
and pressure in a linearized test of the full cosmological Euler
equations.

After linearization as in the non-expanding Jeans length case we find
\begin{equation}
   \partial_t^2 \rho_1 + \frac{2 \dot a}{a} \partial_t \rho_1 - [v_s^2
      \partial^2 \rho_1 + \frac{\rho_0 \partial^2 \phi}{a^3}] = 0
\end{equation}
and expanding in spatial fourier modes we find
\begin{equation}
   \ddot \rho_1 + \frac{2 \dot a}{a} \dot \rho_1 + [v_s^2 k^2 -
      \frac{4 \pi G \rho_0}{a^3}]\rho_1 = 0.
\end{equation}
We know from the homogeneous solutions that $v_s \sim
v_{s_0}\frac{t}{t_0}^{-\frac{8}{3}}$, from which we find solutions 
\begin{equation}
   \rho = \rho_0 (1 + \epsilon t^{-\frac{1}{6}} J_{-\frac{5}{2}}(d)
      \cos{kx})
\end{equation}
\begin{equation}
   p = p_0 (1 + \epsilon \gamma t^{-\frac{1}{6}} J_{-\frac{5}{2}}(d)
      \cos{kx})
\end{equation}
\begin{equation}
   v = -\epsilon \frac{t^{-\frac{7}{6}}}{k} [\frac{d}{3}
      J_{-\frac{3}{2}}(d) + \frac{2}{3} J_{-\frac{5}{2}}(d)]sin(kx)
\end{equation}
where $d \equiv 3 v_{s_0} k t_0^{\frac{4}{3}} t^{-\frac{1}{3}}$, and
$v_{s_0}$ is the initial speed of sound, $t_0$ is the initial time,
$k$ is the wavenumber and $J_i$ is the i'th Bessel function.

The first fourier mode for the solution with a $64$ grid spacing
linear scale is plotted in figures 5 and 6. Note that since the
collapse is now a power law, as opposed to exponential, the density is
correctly resolved much further into the future than for the
non-expanding case.

\subsection{Convergence Test}

To test convergence of solutions on the scale of a few grid spacings,
We compare the Jeans length solution at a given scale with the exact
analytical result. We run the test on grids of $4$, $8$, $16$, $32$ and
$64$ grid spacings. The results are plotted in figure 7.

The best resolution is on the largest scales. At $8$ grid spacings,
that is $4$ grid zones to resolve one bump in the sine function, the
solution is still good to about $85$ percent by the end of the run,
but for $4$ grid spacings the comparison is much poorer. This
indicates that we can only trust the code to resolve features on the
$3$ or $4$ grid spacing level. On scales larger than $32$ grid
spacings the code matches the analytical result to better than $99\%$.
As expected from the finite resolution, the solution is poorer as the
density becomes concentrated in one cell and fails to be resolved by
the grid.

\subsection{Self-Similarity Test}

For one dimensional problems in a flat (no spatial curvature)
background, the fluid should approach a self-similar solution due to
the lack of a characteristic length scale. To test this, we set up an
initial density perturbation along a symmetry axis for the planar and
spherically symmetric cases and watch the evolution. For the initial
perturbation, we simply put an overdensity in one gridzone, then let
the fluid evolve. For these initial conditions, the boundary condition
for the self-similar solution is $\rho$ equal to the background
density, $v$ equal to zero, and $p$ equal to the background pressure
as scaled with the expansion.

To identify the solution we look at the density. We take the
self-similar scale to be identified by the point where the density
goes to the average background density, thus identifying the boundary
of the self-similar solution. We could alternatively have taken a
particular feature of the solution as the self-similar scale. In the
spherically symmetric case, we show results in figures 8a - 8e. The
boundary point recedes from the symmetry axis, and the solution is
well resolved over about 20 expansion times.

For the figures, we have normalized the density maximum to a constant.

and the spherically symmetric case is shown in figures 8a - 8e.

\subsection{Testing the PIC code}

The particle-in-cell method has been extensively tested in the
literature, and its drawbacks and strengths are well known (Hockney \&
Eastwood 1988, Efstathiou \& Eastwood 1981). Thus, we have only tested
our PIC code to the extent necessary to ensure that it is functioning
correctly.

We have made three tests of the PIC code to ensure that there are no
bugs and the constants and parameters are correct. First, we tested
a linear gravitational perturbation (Zel'dovich pancake) in a
non-expanding background to ensure that the growth was exponential.
Next, we tested a homogeneous velocity distribution and made sure that
the velocity fell off as $a^{-2}$. And, finally, we tested a linear
gravitational perturbation in an expanding background. In all cases,
the numerical solution is good to within a small fraction of a percent
of the analytical solution.

\subsection{Testing the Combined PPM \& PIC Code}

We have tested the combined PPM-PIC code less extensively than the
individual codes. We have relied primarily on results from energy
conservation and self-similarity of solutions to test the code. Total
energy is conserved by the code to a fraction of a percent.

For our coordinates, the total energy of the fluid is
\begin{equation}
   K_t + U_t + W_t + \sum_{x = t_{init}}^{t}\dot a W_x = K_0 + U_0 +
      W_0 
\end{equation}
where $K_t$, $U_t$ and $W_t$ are the kinetic, internal and potential
energies at time $t$, $a$ is the scale factor and a $0$ subscript
indicates the initial value of one of the energies (see Peebles (1980)
for a derivation of the energy conservation equation in comoving
coordinates and Hockney and Eastwood (1988) for the above form of the
energy conservation equation).

We measure energy conservation by dividing the difference in total
energy from the beginning to the end of the simulation by the
difference in potential energy from the beginning to the end of the
simulation: $\frac{\Delta E}{\Delta W}$. Or, if the potential hasn't
changed by much (as is usually the case early on in a simulation), we
divide the difference in total energy by the total initial energy:
$\frac{\Delta E}{E_0}$. Typically, by the end of a long simulation,
the former measure is smallest, indicating that the potential energy
has changed by a few orders of magnitude more than the total energy,
but that the total energy has remained constant to within less than a 
percent of the change of the potential.

We tested self-similarity of the combined code in the planar symmetric
case. Instead of introducing a single gridzone overdensity, as we did
to test the PPM code alone, we start with an initially flat density
distribution and add a velocity kick inward to the symmetry plane from
two fluid populations on either side of the symmetry plane causing two
incoming streams to flow toward the symmetry axis. This initial
condition was chosen due to its relevance to the cosmic string model
of large scale structure formation (being investigated by A.S.).

The densities of combined (dash-dot), dark (dash) and baryonic fluids
(solid), the velocity profile and the energy density are shown in
figures 9a - 9f. In the figures, only the portion of the simulation
where the non-linear feature occurs (approximately 20 grid spacings)
is shown.

The initial redshift for the simulation was $z = 800$. The baryonic
temperature at this time is set equal to the background radiation
temperature. We can see physically what is happening: the
collisionless matter forms two streams one on each side of the
symmetry axis. The streams flow through each other then fall back
toward the symmetry plane creating dark matter overdensities to either
side of the symmetry plane. Eventually, secondary and tertiary peaks
form as the dark matter flows back and forth about the symmetry axis.
These peaks are poorly resolved on the scale of the simulation and
form a single overdense lump at the symmetry axis. 

The collisional fluid collides and creates an overdensity at the plane
of symmetry with a shock which is stationary in comoving coordinates
(but outward moving in physical coordinates). Initially, due to the
higher temperature and consequent high sound speed the overdensity
broadens. Then as the sound speed decreases due to the expansion, the
fluid clumps closer to the symmetry axis. This result has been
anticipated in one dimension by Hara and Miyoshi (1987).  The solution
approaches self-similarity at a redshift of about $z = 200$ but
self-similarity fails relatively quickly since the matter streams
deplete the matter on either side of the symmetry axis by a redshift
of $z = 100$ and a bound system remains about the symmetry axis. The
fact that the matter is depleted is not a physical result here. It is
just a reflection of the fact that the boundary conditions for the
code do not allow for fluid inflow from the boundaries.

\section{Conclusions}

We have presented an Eulerian PPM/PIC code for simulation of
cosmological hydrodynamics incorporating gravitational and expansion
source terms in the Euler equations and a collisionless dark matter
component coupled via gravity to the collisional baryonic fluid. We
have made what we consider the minimal changes which keep the higher
order accuracy of the method intact.

We have found that the code is accurate to a length scale of $\sim
3-4$ grid spacings, as shown by convergence testing; and one
dimensional solutions approach self-similarity, indicating that the
physics of the equations is well modeled.

Our incorporation of a non-gamma-law Riemann solver adds extra
potential for simulation of interesting astrophysical situations, but
has the disadvantage of taking up extra memory and run time. 

The largest disadvantage of our code is that the resolution is limited
by the memory size of the MP-2. Due to this constraint, we are limited
to calculations of $128^3$ elements. Porting to a more powerful
machine with a larger memory or incorporating adaptive techniques
would provide the capability to do higher resolution simulations.

\newpage

\begin{center}
\begin{tabular}{||l|l||} 
\hline
\multicolumn{2}{||c||}{\bf TABLE 1} \\ \hline
\multicolumn{2}{||l||}{\it Timing Breakdown of PPM Code} \\ \hline
Operation             & Seconds \\ \hline
Flux Calculation X    & 8       \\ \hline
Flux Calculation Y    & 8       \\ \hline
Swap Z to Processors  & 8       \\ \hline
Flux Calculation Z    & 8       \\ \hline
Swap Z Back to Memory & 8       \\ \hline
Poisson Solver        & 2.2     \\ \hline\hline
Total                 & 42.2    \\ \hline
\end{tabular}
\end{center}
\vspace{0.25truein}

\newpage

\begin{center}
\begin{tabular}{||l|l||} 
\hline
\multicolumn{2}{||c||}{\bf TABLE 2} \\ \hline
\multicolumn{2}{||l||}{\it Timing Breakdown of PIC Code} \\ \hline
Operation            & Seconds \\ \hline
Deposit              & 1.4     \\ \hline
Poisson Solver       & 2.2     \\ \hline
Force Interpolation  & 3.3     \\ \hline\hline
Total                & 6.9     \\ \hline
\end{tabular}
\end{center}
\vspace{0.25truein}

\rm
\newpage

\Large\bf
References\\
\rm\normalsize
Bryan, G., Norman. M., Stone, J., Cen, R., \& Ostriker, P. 1995, 89, 149\\
Cen, R. 1992, ApJS, 78, 341\\
Colella, P. \& Woodward, P. 1984, J. Comp. Phys., 54, 174\\
Efstanthiou, G. \& Eastwood, J. 1981, MNRAS, 194, 503\\
Fryxell, B., \& Taam, R. 1989, ApJ, 339, 297\\
Fryxell, B., Muller, E., \& Arnett, D. 1991, ApJ, 367, 619\\
Fryxell, B., Zylstra, G., \& Melia, F. 1992, ApJ, 396, L27\\
Hara, T. \& Miyoshi, S. 1987, Prog. Theor. Phys., 77, 1152\\
Hockney, R. \& Eastwood, J. 1988, Computer Simulations using\\
\indent
Particles (New York: Adam Hilger).\\
MacNeice, P., Mobarry, C., \& Olson, K. 1995, submitted SIAM\\
Peebles, P. J. E. 1980, The Large-Scale Structure of the Universe\\
\indent
(Princeton: Princeton Univ. Press).\\
Woodward, P. \& Colella, P. 1984, J. Comp. Phys., 54, 115\\

\newpage
\Large\bf
Figure Captions\\
\normalsize\rm

Fig. 1 - Energy decrease in homogeneously expanding universe. The
numerical results are plotted against the analytical solution (solid
line).\\
\\
Fig. 2 - Velocity decrease in homogeneously expanding universe. The
numerical results are plotted against the analytical solution (solid
line).\\
\\
Fig. 3 - The solution for initial pressure $p > p_{crit}$ for the
non-expanding Jeans length test. The amplitude of the first fourier
mode is plotted against the analytical solution (solid line).\\

Fig. 4 - The solution for initial pressure $p < p_{crit}$ for the
non-expanding Jeans length test. The amplitude of the first fourier
mode is plotted against the analytical solution (solid line).\\
\\
Fig. 5 - The solution for initial pressure $p = 500 p_{crit}$ for the
expanding Jeans length test. The numerical results are plotted
against the analytical solution (solid line).\\
\\
Fig. 6 - The solution for initial pressure $p = 0.9 p_{crit}$ for the
expanding Jeans length test. The numerical results are plotted
against the analytical solution (solid line).\\
\\
Fig. 7 - The Jeans length test densities computed on size $32$, $16$,
$8$ and $4$ grids are plotted as a fraction of the analytical
solution. The size $32$ grid result is uppermost. It declines from
exact match of the analytical result to about $95$ percent of the
analytical result by the end of the test. Size $16$ and $8$ grids
perform successively worse, and the lowermost (size $4$) result
obviously completely fails to reproduce correct densities.\\
\\
Fig. 8a - Approach to self-similarity after 200 timesteps (spherical).\\
Fig. 8b - Approach to self-similarity after 400 timesteps (spherical).\\
Fig. 8c - Approach to self-similarity after 600 timesteps (spherical).\\
Fig. 8d - Approach to self-similarity after 800 timesteps (spherical).\\
Fig. 8e - Approach to self-similarity after 1000 timesteps (spherical).\\
\\
Fig. 9a - Planar symmetric dual stream solution with combined code.\\
Fig. 9b - Planar symmetric dual stream solution with combined code.\\
Fig. 9c - Planar symmetric dual stream solution with combined code.\\
Fig. 9d - Planar symmetric dual stream solution with combined code.\\
Fig. 9e - Planar symmetric dual stream solution with combined code.\\
Fig. 9f - Planar symmetric dual stream solution with combined code.\\
\\
\newpage
\psfig{figure=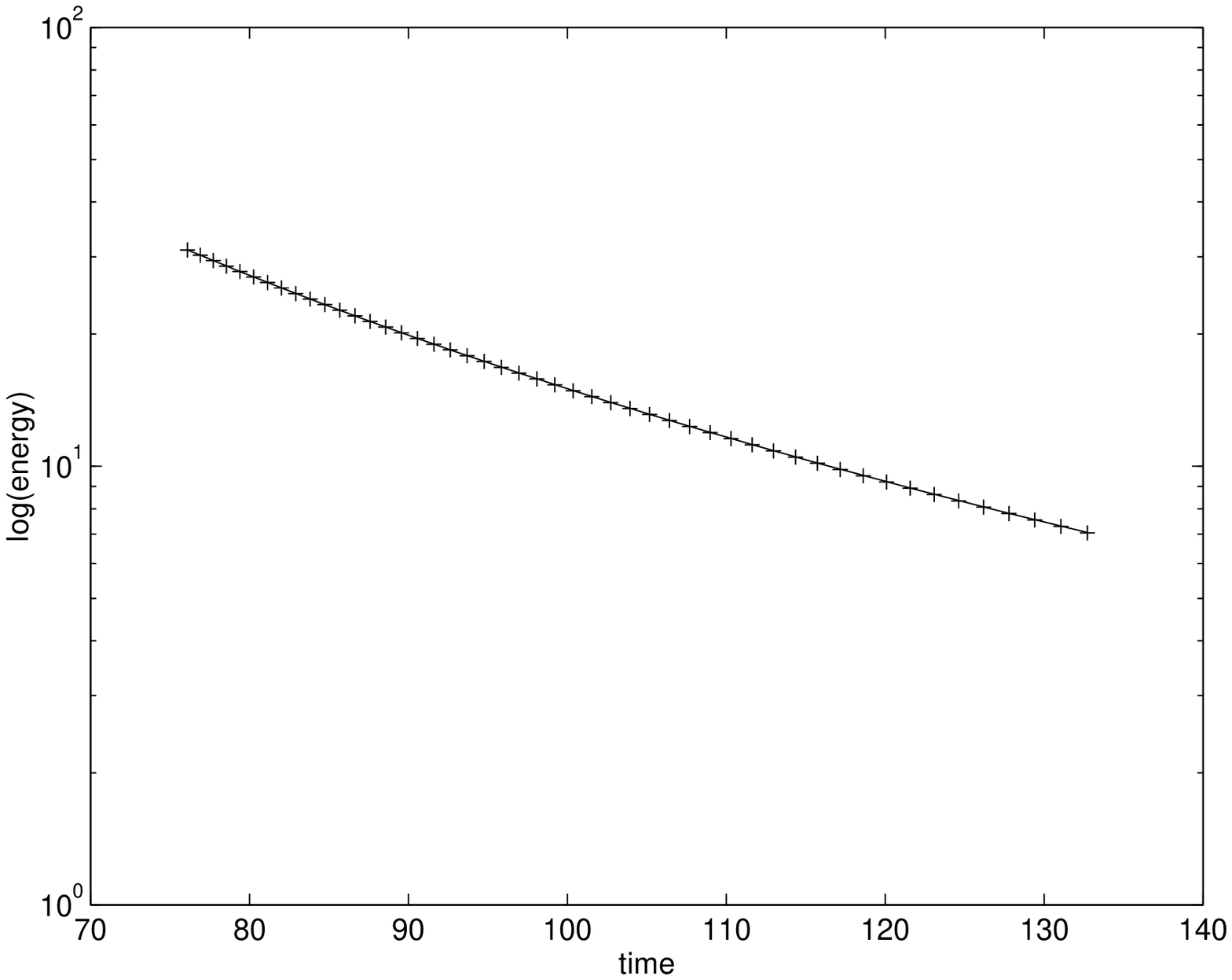,height=7.0in,width=5.0in}
\psfig{figure=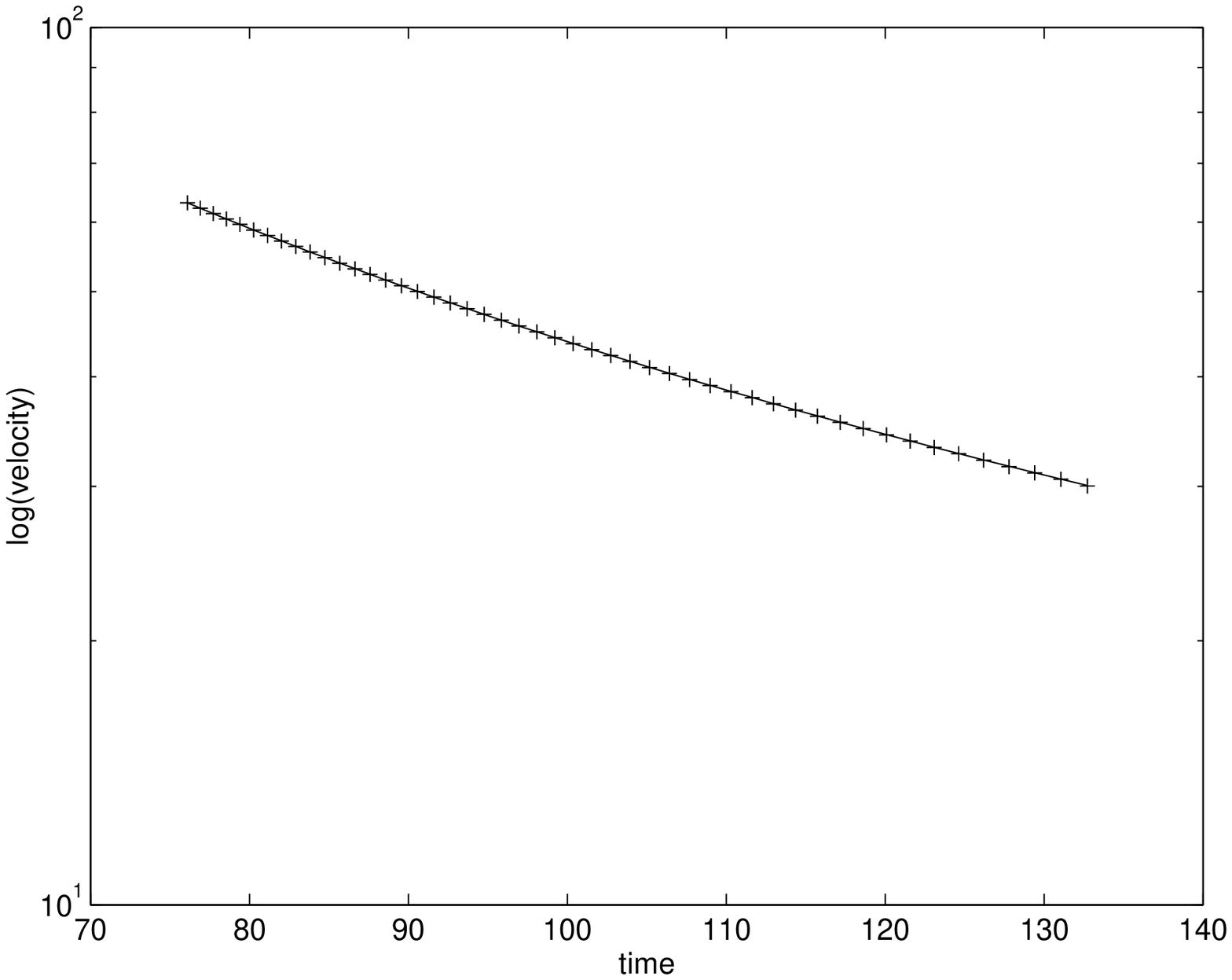,height=7.0in,width=5.0in}
\psfig{figure=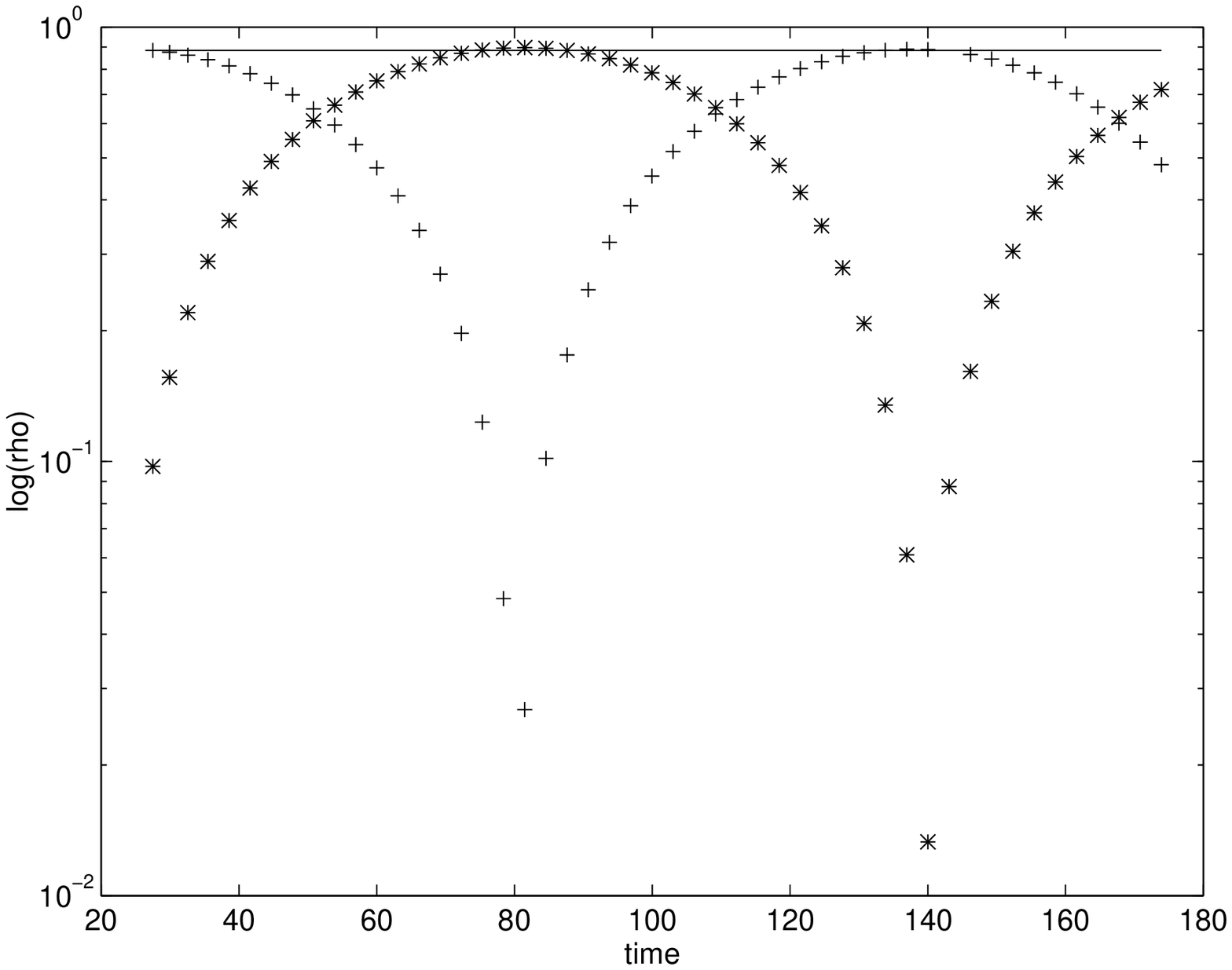,height=7.0in,width=5.0in}
\psfig{figure=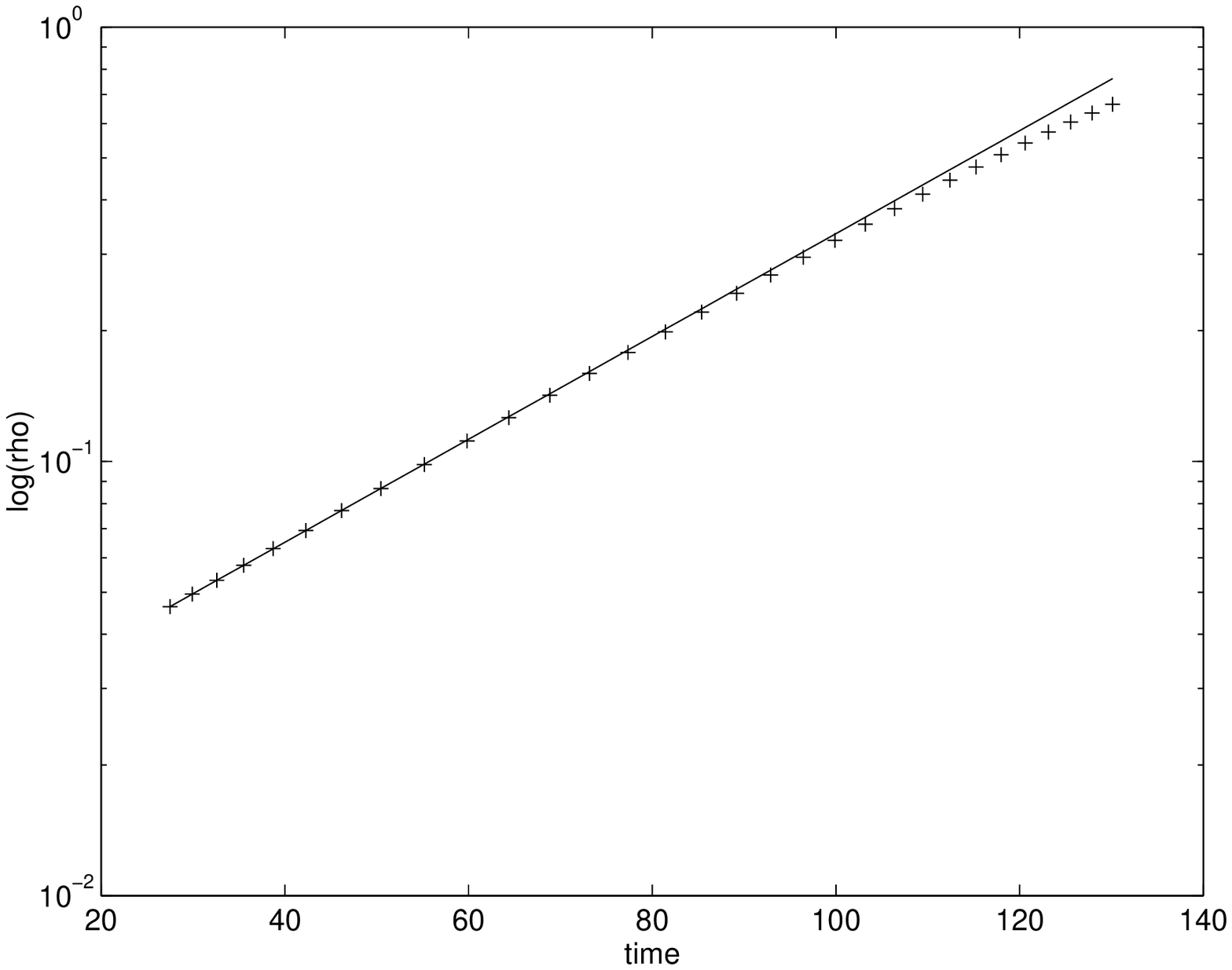,height=7.0in,width=5.0in}
\psfig{figure=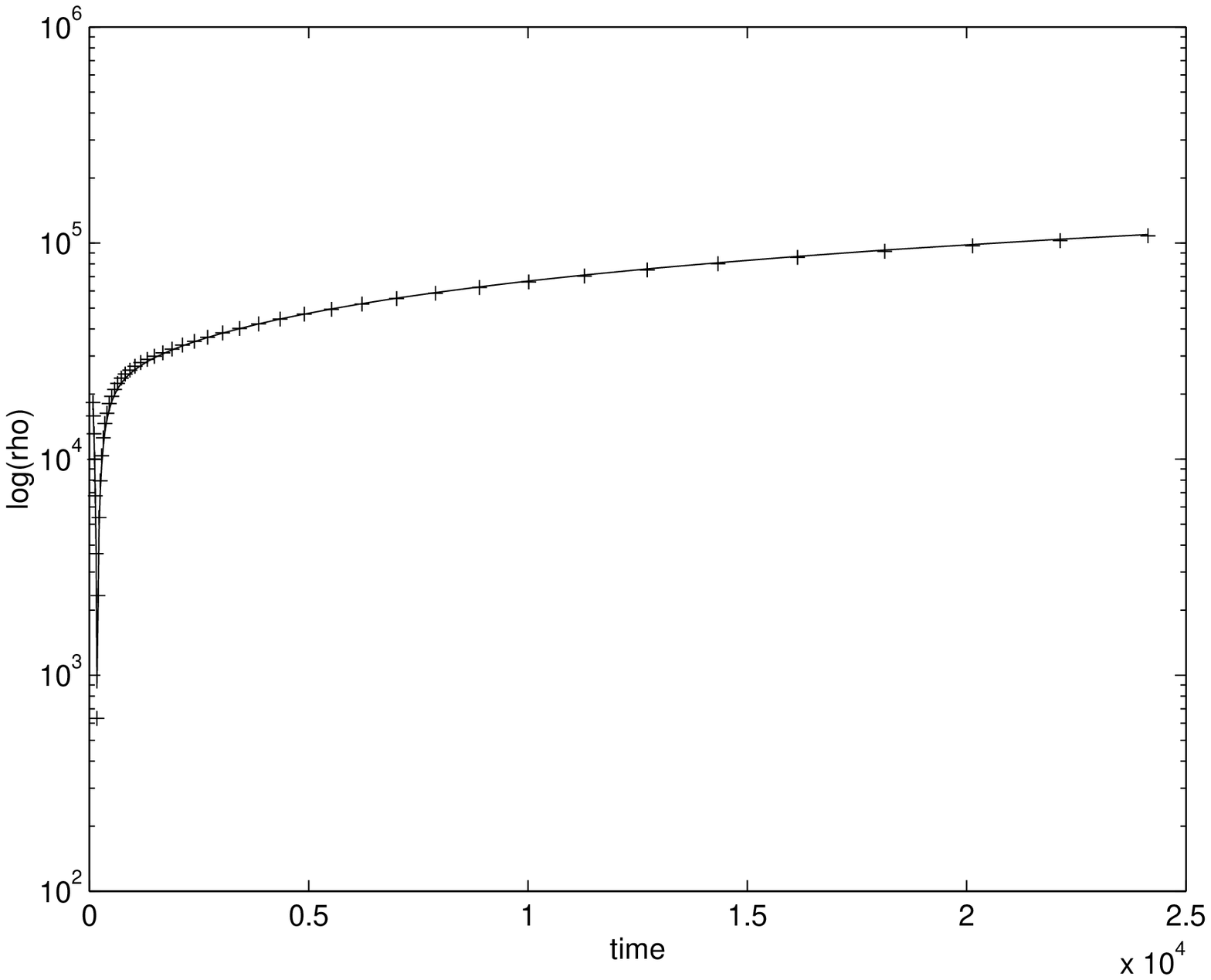,height=7.0in,width=5.0in}
\psfig{figure=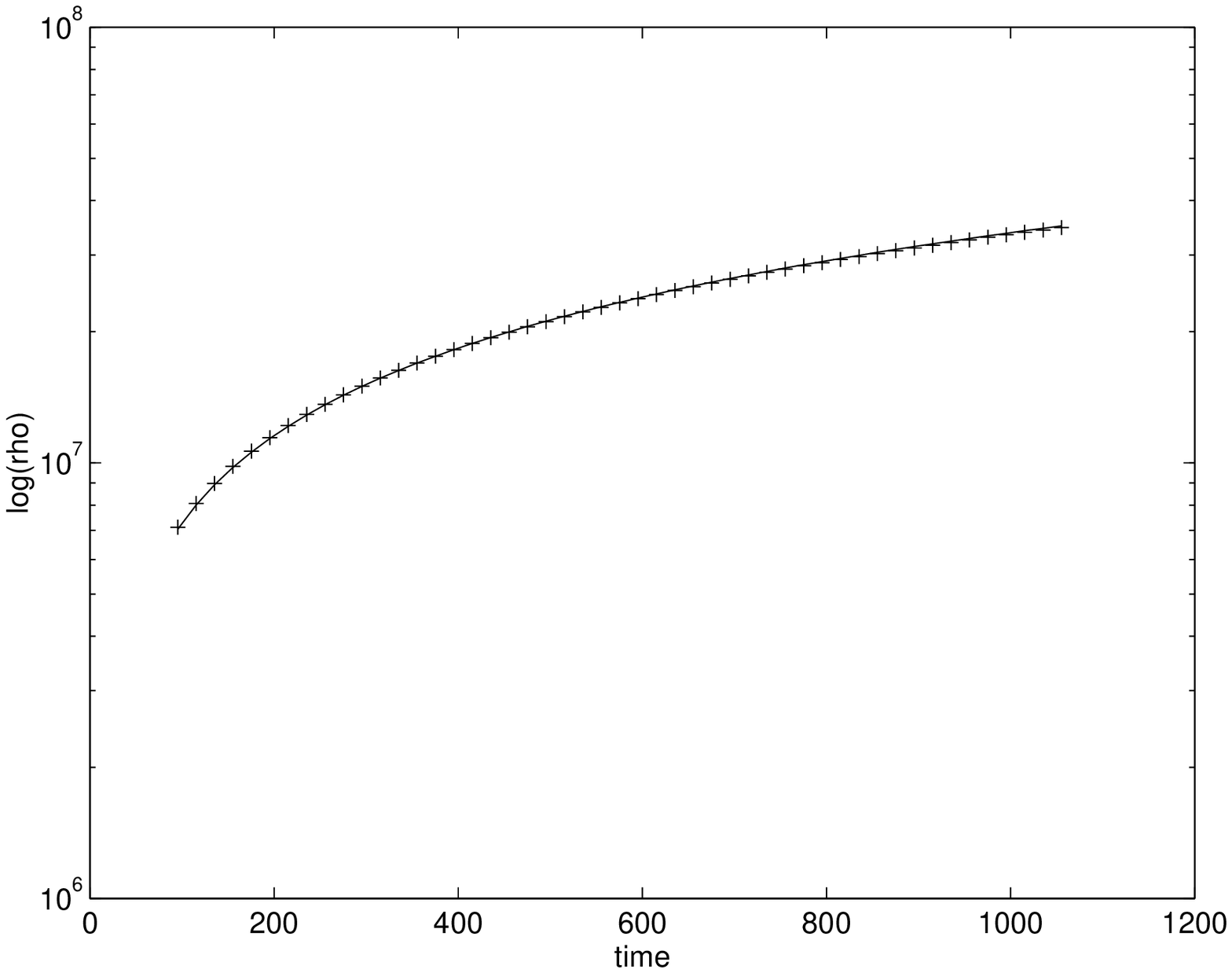,height=7.0in,width=5.0in}
\psfig{figure=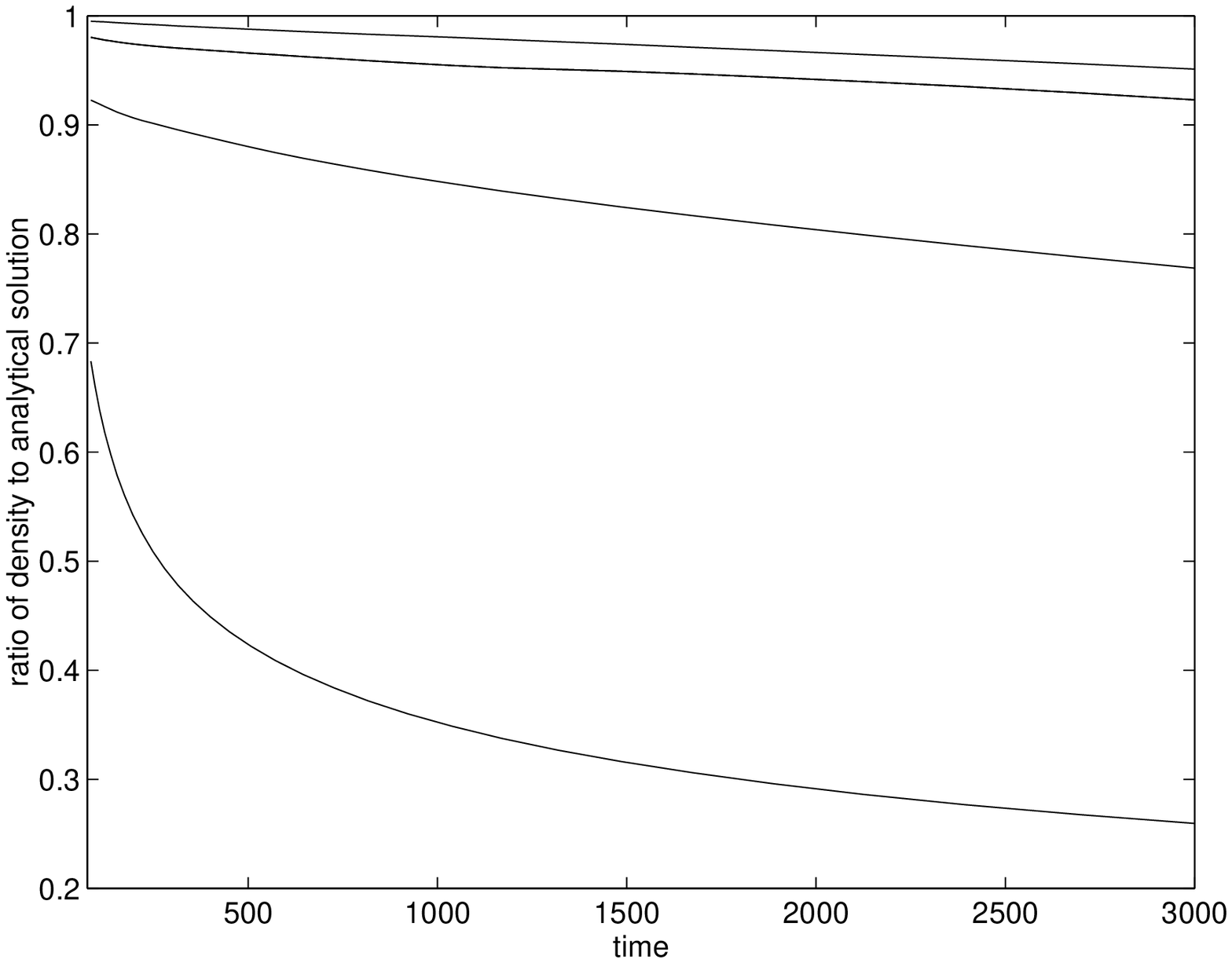,height=7.0in,width=5.0in}
\psfig{figure=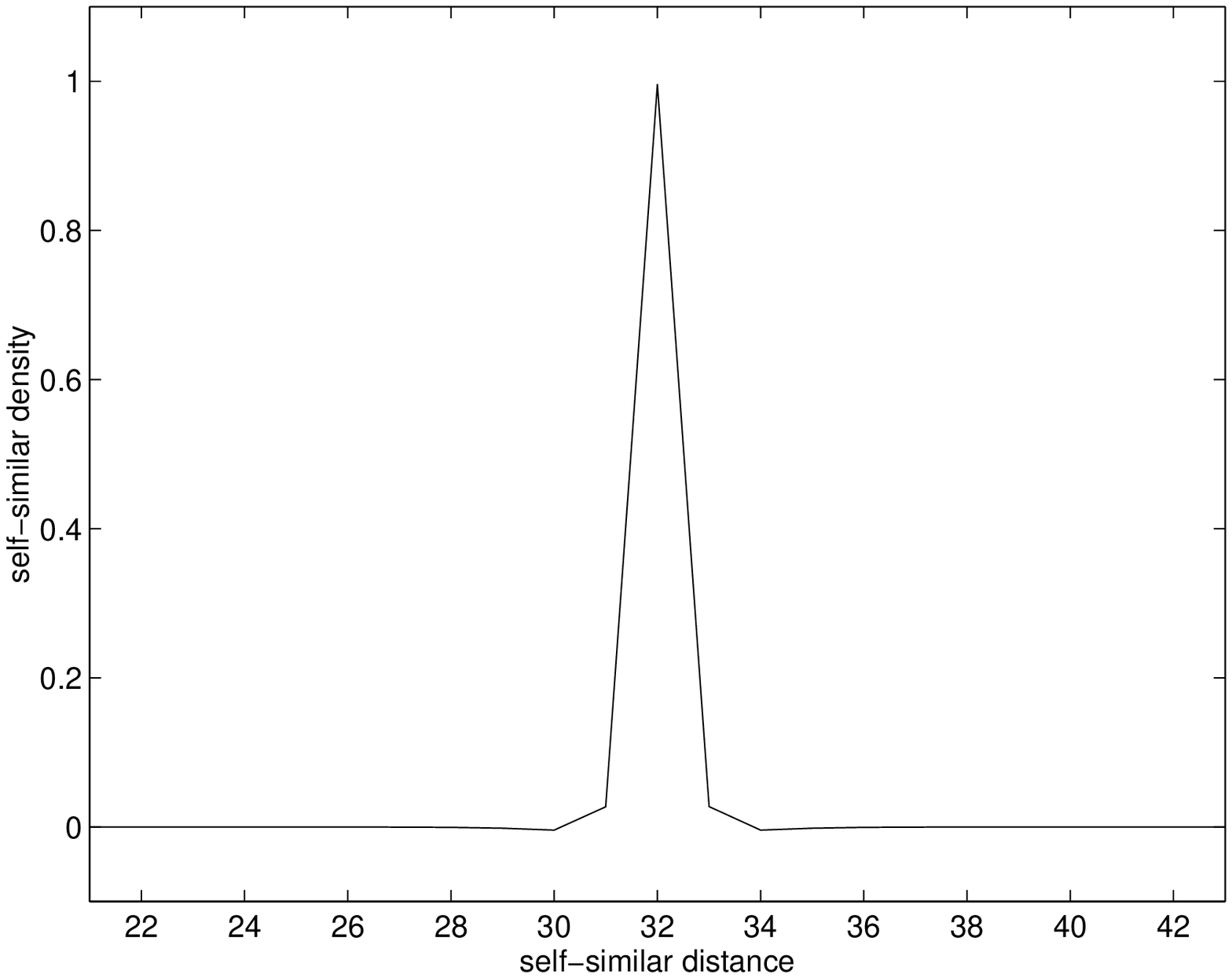,height=7.0in,width=5.0in}
\psfig{figure=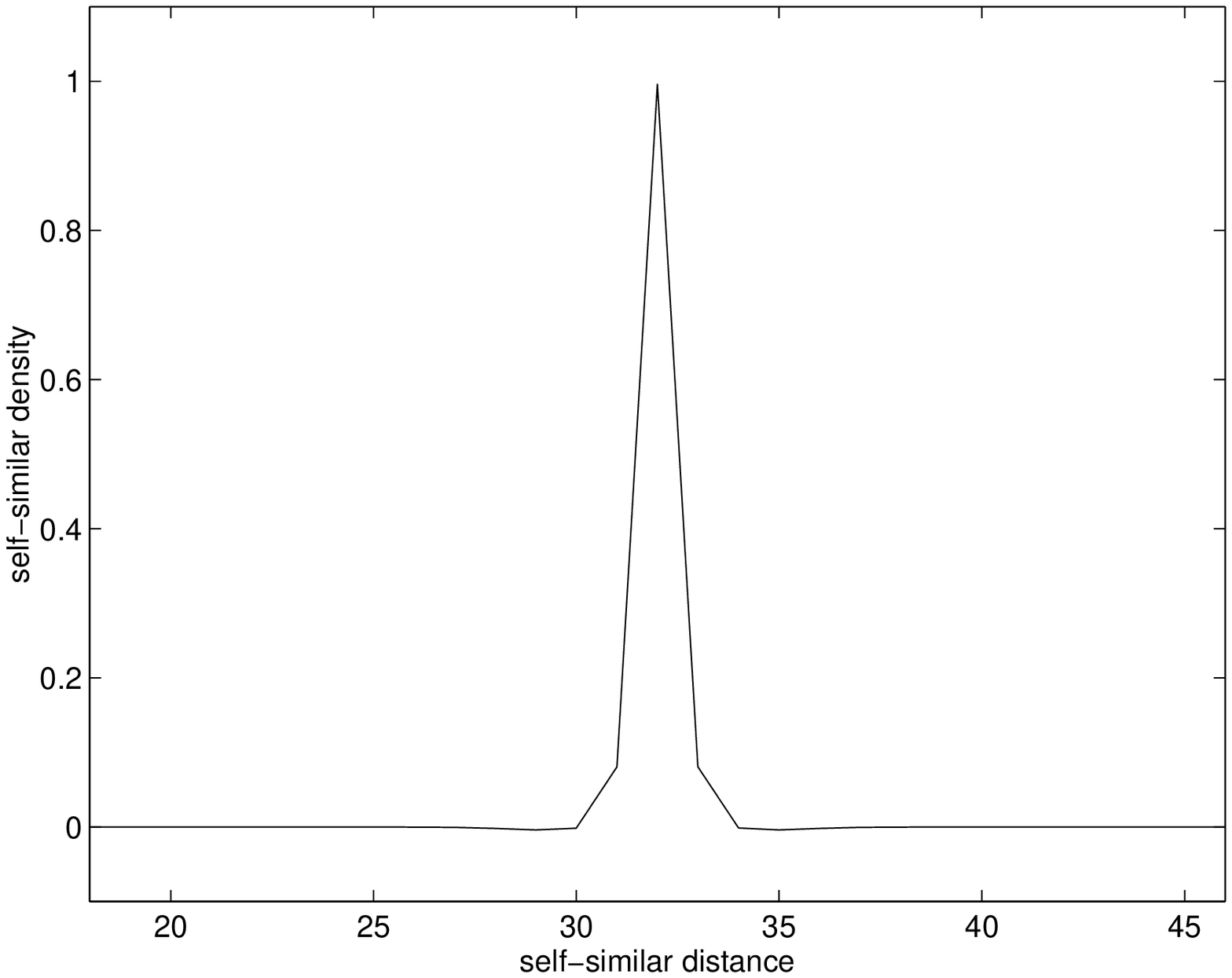,height=7.0in,width=5.0in}
\psfig{figure=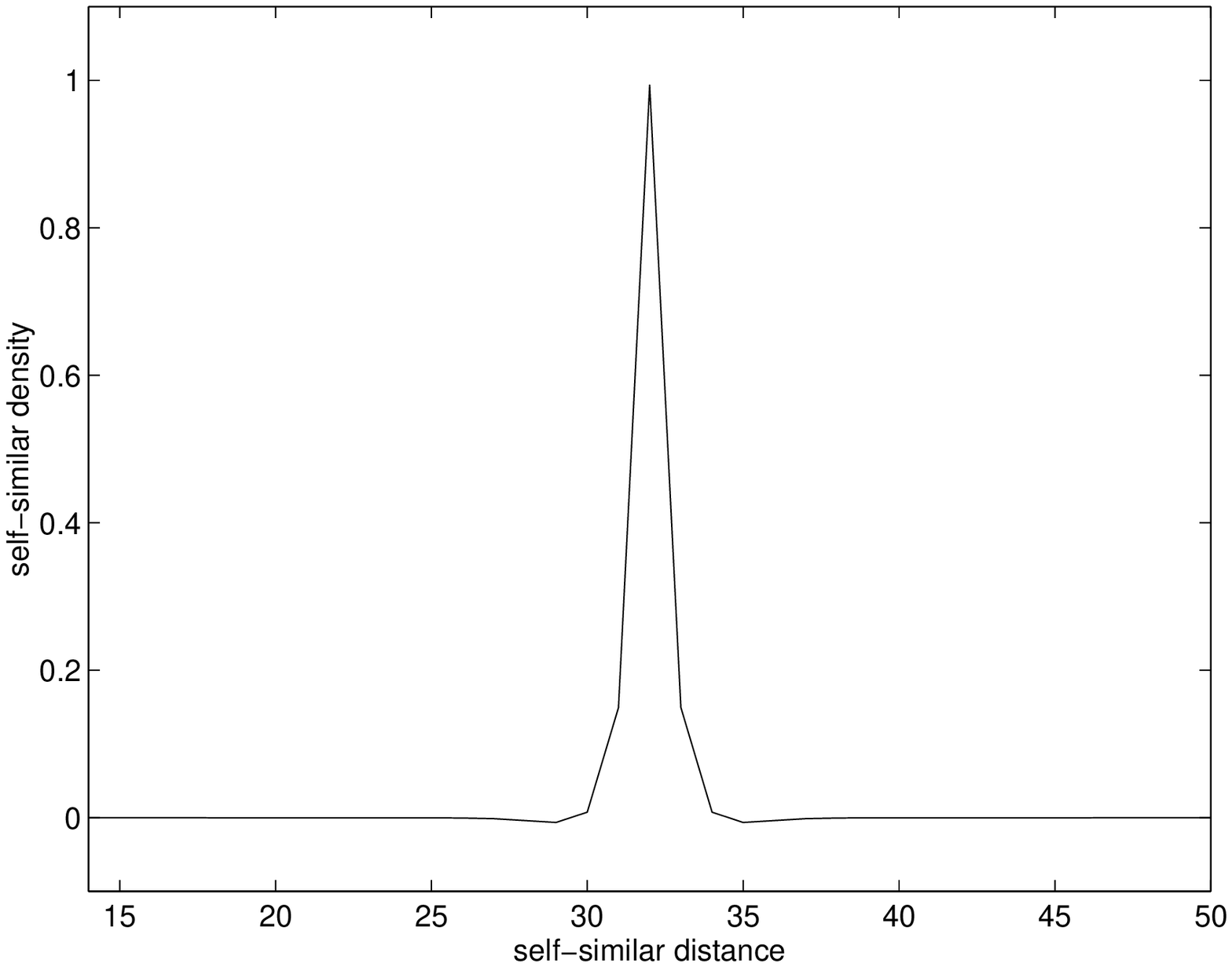,height=7.0in,width=5.0in}
\psfig{figure=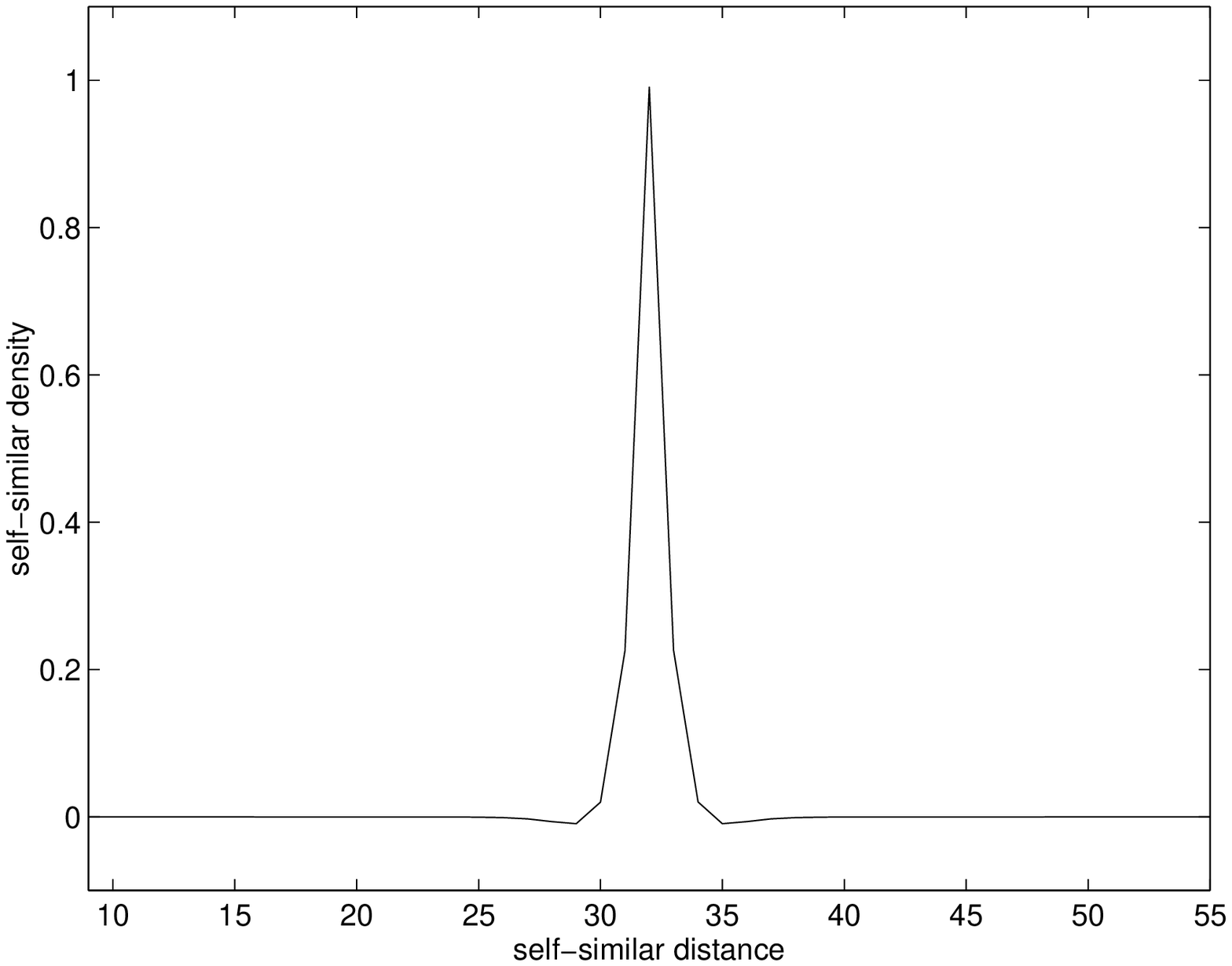,height=7.0in,width=5.0in}
\psfig{figure=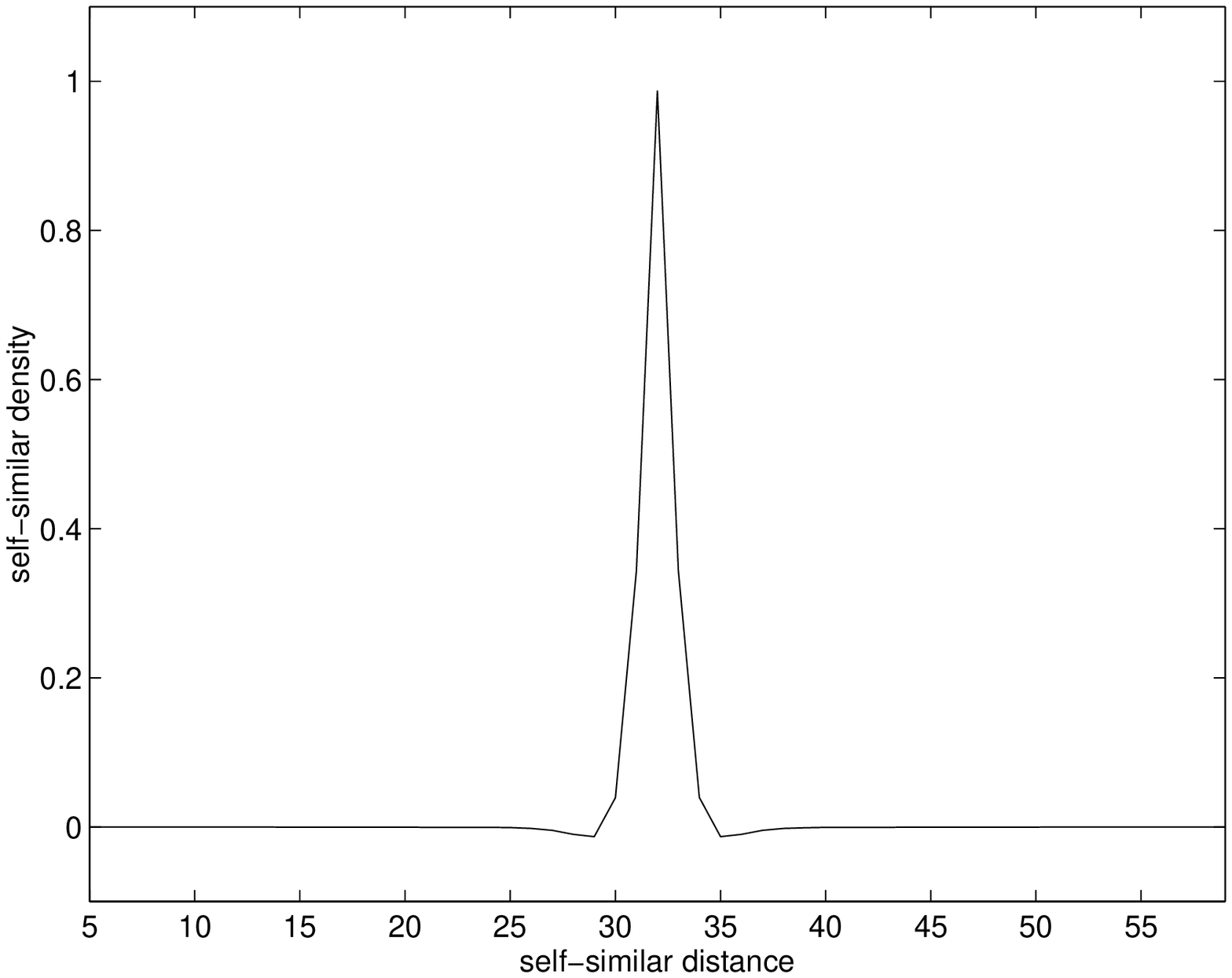,height=7.0in,width=5.0in}
\psfig{figure=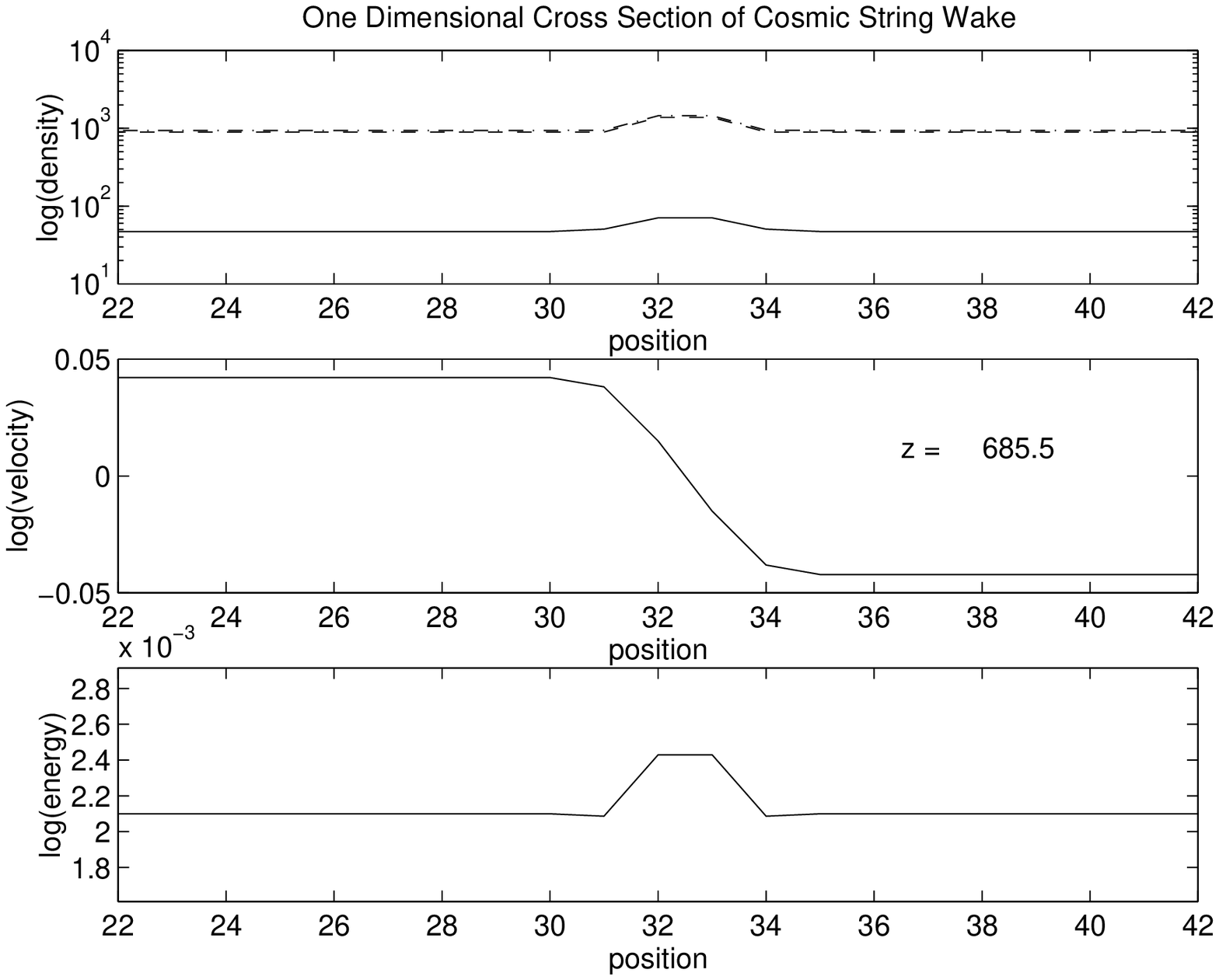,height=7.0in,width=5.0in}
\psfig{figure=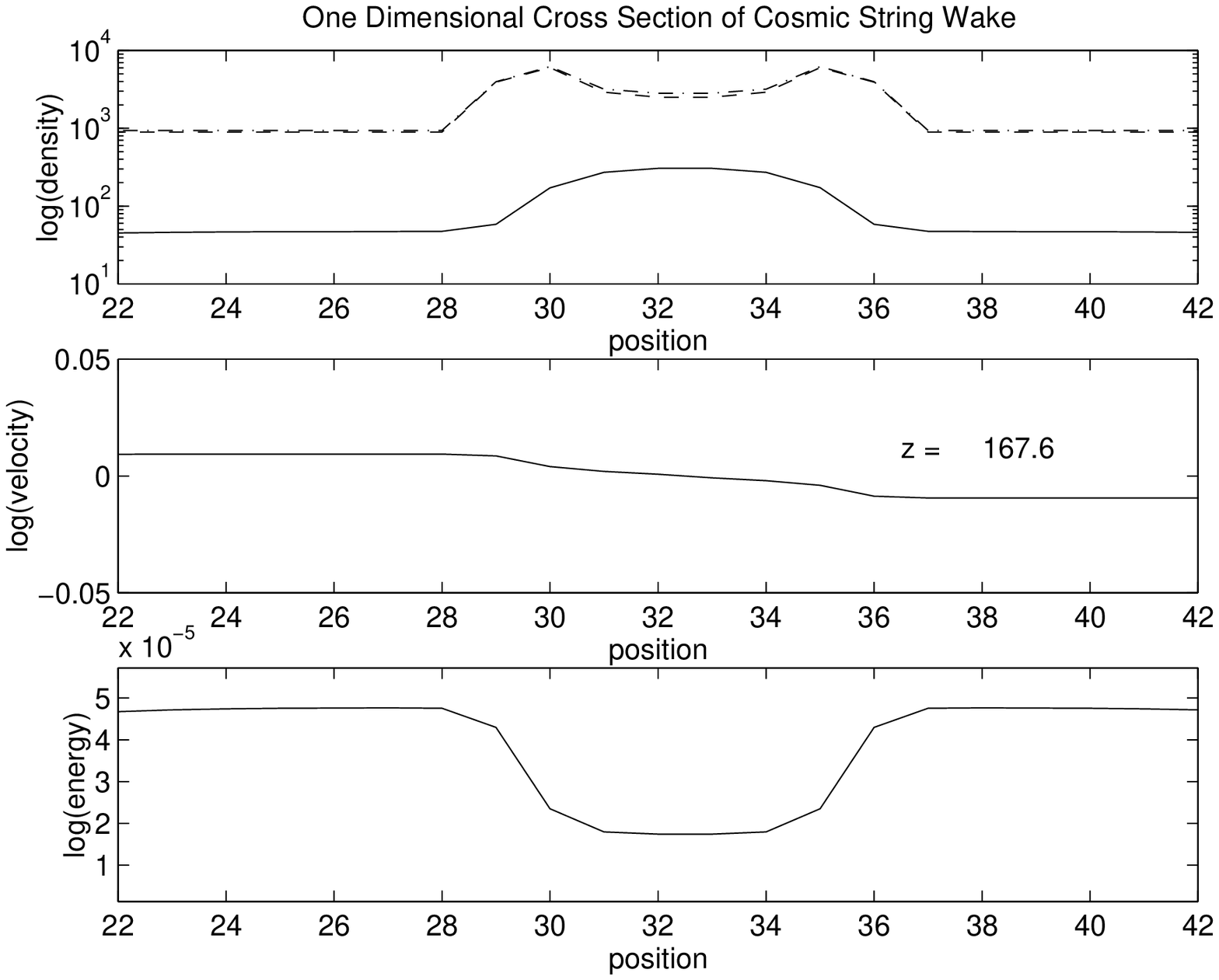,height=7.0in,width=5.0in}
\psfig{figure=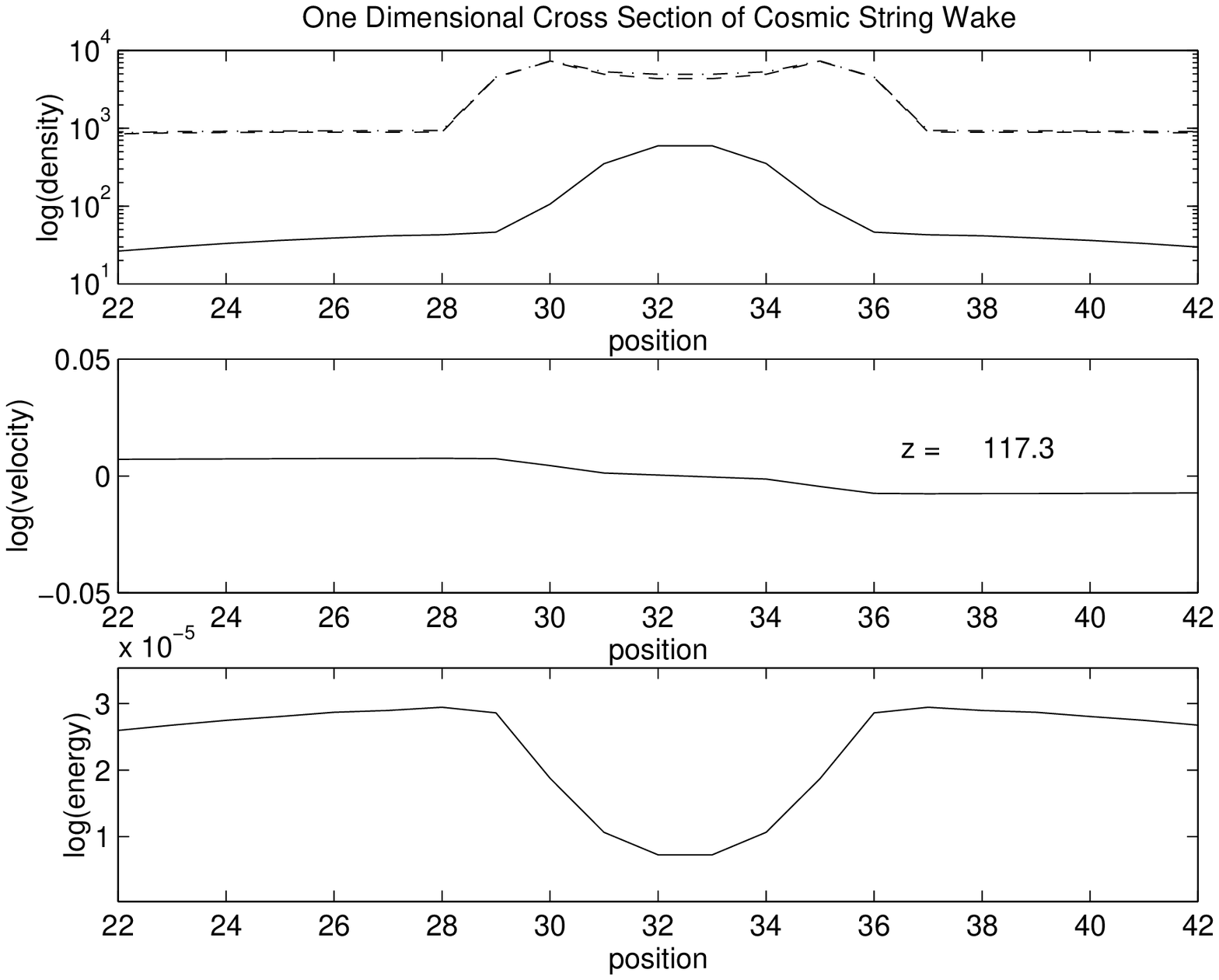,height=7.0in,width=5.0in}
\psfig{figure=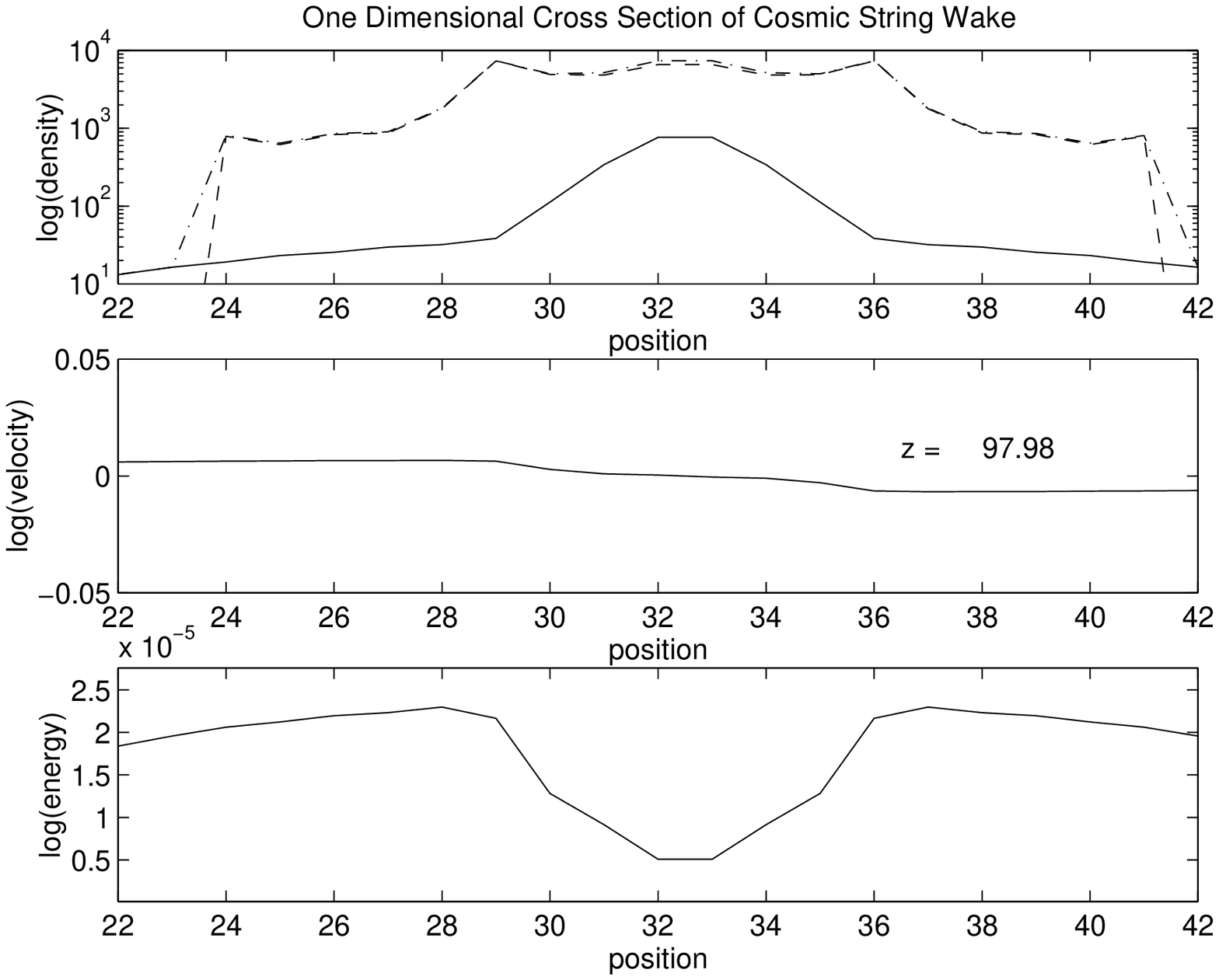,height=7.0in,width=5.0in}
\psfig{figure=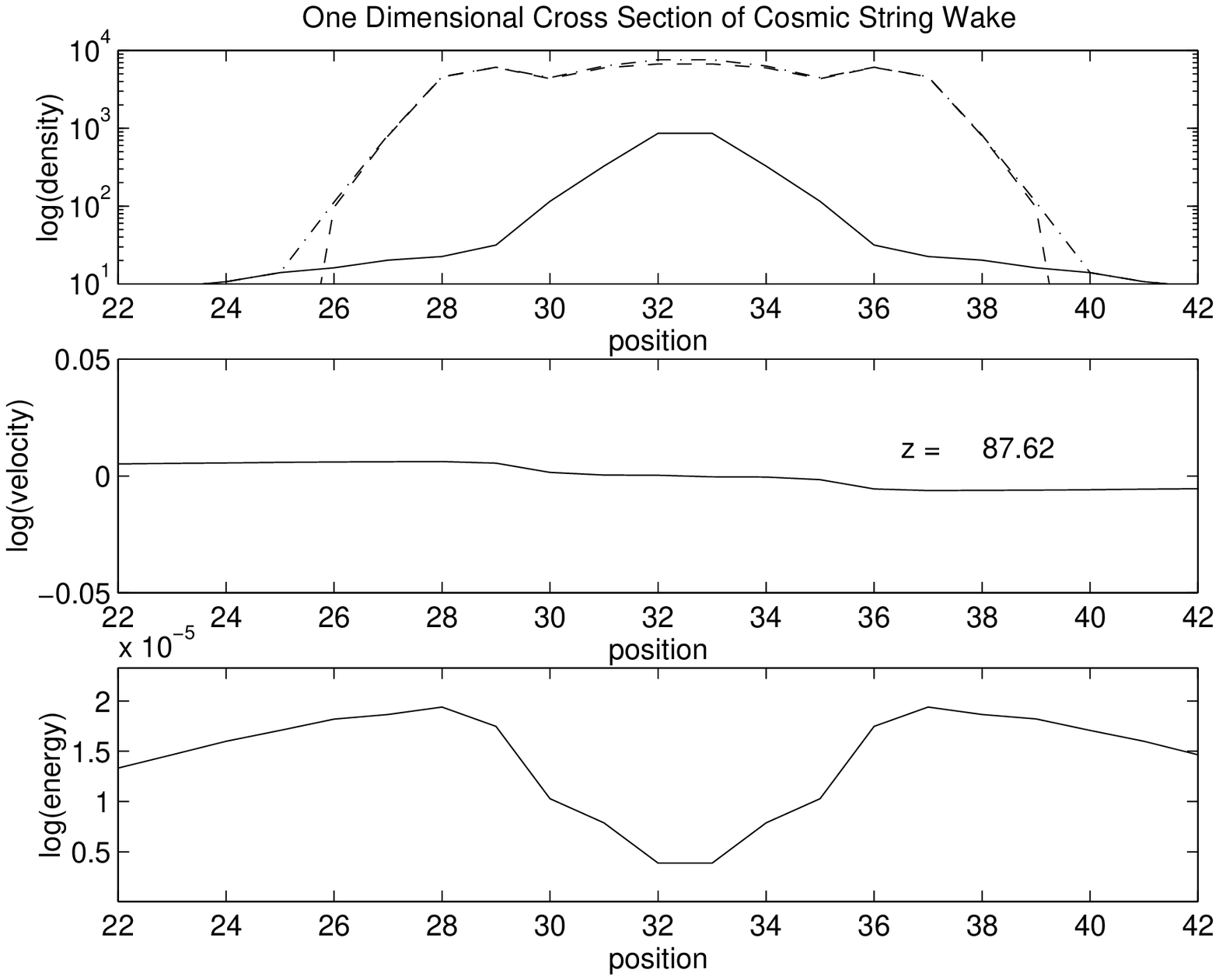,height=7.0in,width=5.0in}
\psfig{figure=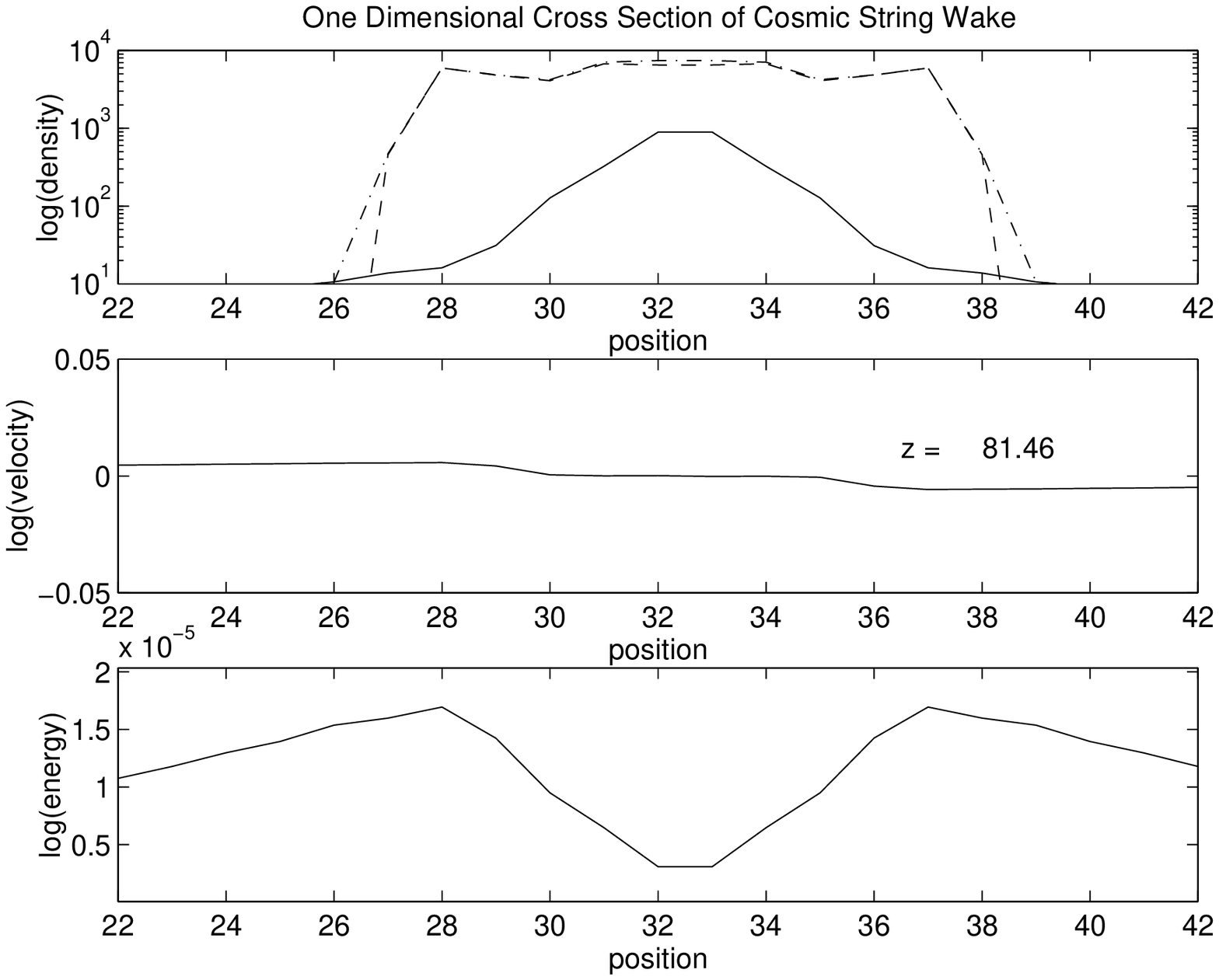,height=7.0in,width=5.0in}

\end{document}